\documentclass[manuscript, screen]{acmart}
\usepackage{xcolor}     
\usepackage{colortbl}

\usepackage{makecell}
\usepackage{booktabs, tabularx, array, enumitem}
\usepackage{caption}
\usepackage{arydshln}
\usepackage{multirow}

\newcolumntype{L}[1]{>{\raggedright\arraybackslash}p{#1}} %
\newcolumntype{Y}{>{\raggedright\arraybackslash}X}
\AtBeginDocument{%
  }

\setcopyright{acmlicensed}
\copyrightyear{2018}
\acmYear{2018}
\acmDOI{XXXXXXX.XXXXXXX}
\acmConference[Conference acronym 'XX]{Make sure to enter the correct
  conference title from your rights confirmation email}{June 03--05,
  2018}{Woodstock, NY}
\acmISBN{978-1-4503-XXXX-X/2018/06}

\begin{document}

\title[Scaffolding Metacognition in Programming Education]{Scaffolding Metacognition in Programming Education: Understanding
Student–AI Interactions and Design Implications}


\author{Boxuan Ma}
\email{boxuan@artsci.kyushu-u.ac.jp}
\authornote{Corresponding author. Email: boxuan@artsci.kyushu-u.ac.jp}
\affiliation{
  \institution{ Kyushu University}
  \city{Fukuoka}
  \country{Japan}
}

\author{Huiyong Li}
\affiliation{
  \institution{ Kyushu University}
  \city{Fukuoka}
  \country{Japan}
}

\author{Gen Li}
\affiliation{
  \institution{ Kyushu University}
  \city{Fukuoka}
  \country{Japan}
}

\author{Li Chen}
\affiliation{
  \institution{ Osaka Kyoiku University}
  \city{Osaka}
  \country{Japan}
}

\author{Cheng Tang}
\affiliation{
  \institution{ Kyushu University}
  \city{Fukuoka}
  \country{Japan}
}

\author{Yinjie Xie}
\affiliation{
  \institution{ Kyushu University}
  \city{Fukuoka}
  \country{Japan}
}

\author{Chenghao Gu}
\affiliation{
  \institution{ Kyushu University}
  \city{Fukuoka}
  \country{Japan}
}

\author{Atushi Shimada}
\affiliation{
  \institution{ Kyushu University}
  \city{Fukuoka}
  \country{Japan}
}

\author{Shin'ichi Konomi}
\affiliation{
  \institution{ Kyushu University}
  \city{Fukuoka}
  \country{Japan}
}

\renewcommand{\shortauthors}{Ma. et al}


\begin{abstract}

Generative AI tools such as ChatGPT now provide novice programmers with unprecedented access to instant, personalized support. While this holds clear promise, their influence on students’ metacognitive processes remains underexplored. Existing work has largely focused on correctness and usability, with limited attention to whether and how students’ use of AI assistants supports or bypasses key metacognitive processes. This study addresses that gap by analyzing student–AI interactions through a metacognitive lens in university-level programming courses. We examined more than 10,000 dialogue logs collected over three years, complemented by surveys of students and educators. Our analysis focused on how prompts and responses aligned with metacognitive phases and strategies. Synthesizing these findings across data sources, we distill design considerations for AI-powered coding assistants that aim to support rather than supplant metacognitive engagement. Our findings provide guidance for developing educational AI tools that strengthen students’ learning processes in programming education. 

\end{abstract}

\ccsdesc[500]{Human-centered computing~Interactive systems and tools}
\ccsdesc[500]{Social and professional topics~Computing education}

\keywords{programming education, generative AI, metacognitive theory, design guidelines}

\maketitle

\section{Introduction}

\begin{figure}
   \centering
    \includegraphics[width=0.9\textwidth]{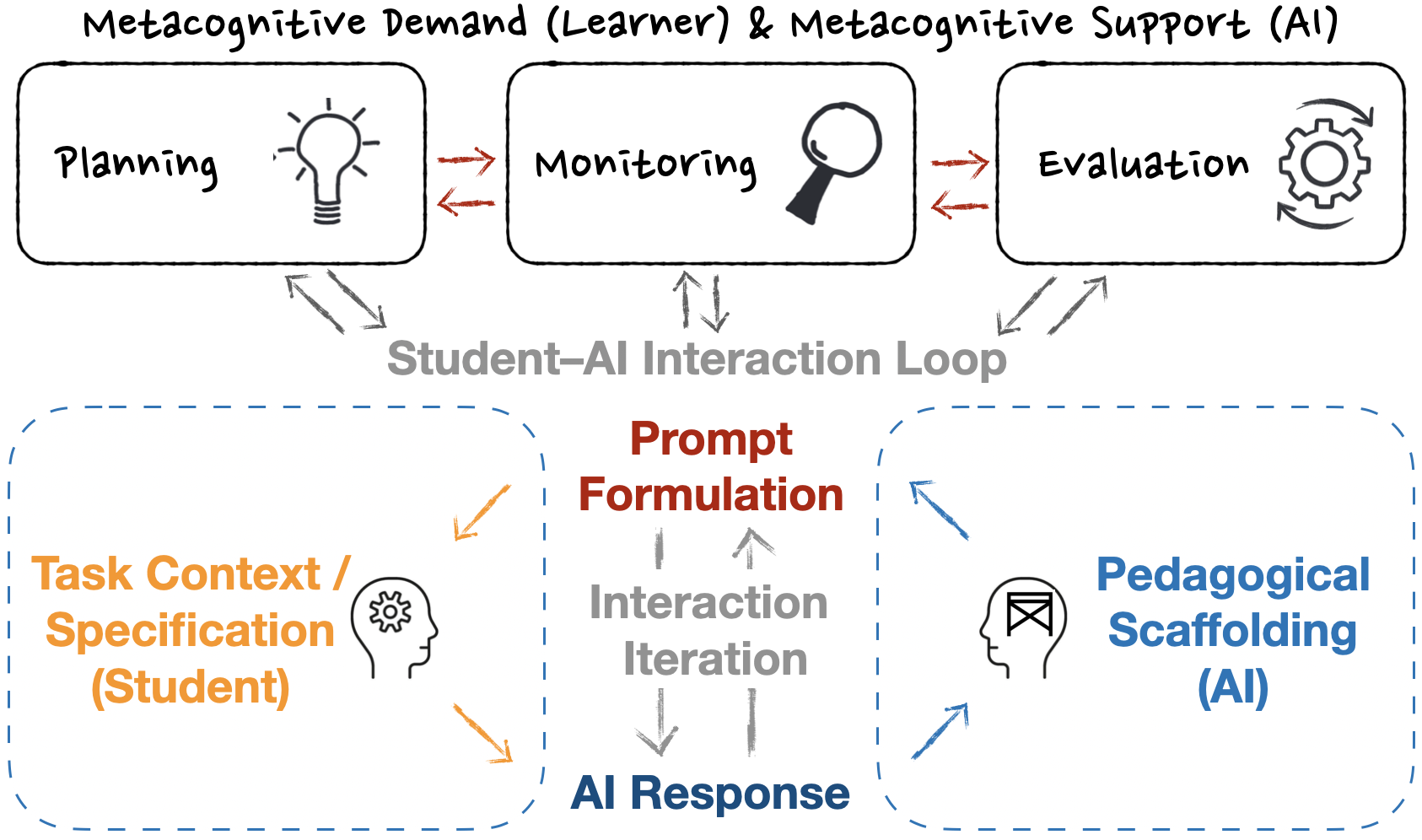}
    \Description{The figure locates metacognitive demand on the learner (task context/specification and prompt formulation) and metacognitive support from the AI (pedagogical scaffolding) within an iterative student–AI loop embedded in the Planning, Monitoring, and Evaluation cycle.}
    \caption{Metacognitive demand (learner) and support (AI) in the student–AI interaction loop.}
    \label{Metacognitive}
\end{figure}

Students learning to program routinely encounter deep uncertainty: they may stare at a blank editor unsure how to begin, struggle to decipher cryptic error messages when code fails, and—even when it runs—doubt whether their solution is efficient, generalizable, or aligned with best practices. These moments often arise precisely when help is hardest to access: office hours may not align with schedules, online forums can feel intimidating, and in crowded classrooms attention tends to go to the most vocal students, leaving feedback uneven and delayed \cite{smith2017office,smith2017my}. In higher-education programming courses, large enrollments, wide variation in prior experience, and limited instructional resources further constrain the timely, individualized support novices need \cite{smith2017office,smith2017my}. Such delays or generic feedback are linked to shallow understanding, weaker retention, limited transfer, and underdeveloped higher-order learning skills \cite{national2018assessing}. Compounding these constraints, many beginners lack a firm grasp of fundamentals (e.g., syntax, control flow, algorithms) and the metacognitive strategies required for planning, monitoring, and evaluation. Addressing this gap calls for scalable, equitable, and learner-centered approaches that deliver guidance at the moment students encounter difficulty.

Large language models (LLMs) and AI-powered coding assistants offer a promising avenue for addressing this need. They can instantly generate explanations, examples, and guidance, providing help precisely when students encounter difficulties \cite{tian2304chatgpt,yilmaz2023augmented}. Recent work highlights how LLMs can augment students’ workflows and make assistance broadly available outside lecture time or office hours \cite{kazemitabaar2024codeaid,sun2024would}. These capabilities make AI a compelling tool for addressing the long-standing gap between when students encounter difficulty and when they receive help. At the same time, this strength of AI tools introduces new pedagogical risks. Because AI can often produce working solutions with minimal prompting \cite{denny2023conversing,finnie-ansley2022robots}, students may offload problem-solving steps to the AI assistants rather than engage in metacognitive processes such as planning, monitoring, and evaluation \cite{anagnostopoulos2023chatgpt,rajala2023call}. Scholars have warned such  “metacognitive laziness” when learners rely on AI to do the thinking for them \cite{darvishi2024impact,fan2025beware}. These concerns extend to academic integrity and to the longer-term development of independent problem solving and metacognitive growth \cite{skjuve2023user,tlili2023if}.

In response, researchers and designers have begun to build constrained, pedagogically minded assistants that avoid simply handing back finished code. Systems such as CodeHelp \cite{liffiton2023codehelp} and CodeAid \cite{kazemitabaar2024codeaid} exemplify interventions that restrict direct answers and instead provide pseudocode, hints, or explanation-first interactions to encourage student reasoning. These guardrails position AI as a partner in thinking rather than a replacement. However, prior work has largely emphasized use patterns and performance outcomes (e.g., correctness, time to solution), with comparatively little attention to metacognitive support. Without designs that align usage with metacognitive phases, such tools risk falling short of their educational potential. We argue that metacognition, understood as the ability to monitor and regulate one’s own thinking, offers a productive lens for improving AI-based assistance. As Figure~\ref{Metacognitive} shows, each turn in the student–AI interaction loop places metacognitive demands on learners and is best met with metacognitive support from the AI. Framing the exchange as a cycle that moves from planning to monitoring to evaluation clarifies when learners should plan, how they should monitor, and how they should evaluate, enabling assistants that strengthen rather than bypass metacognitive skills.

This paper addresses this gap by bringing metacognitive theory into conversation with empirical studies of real-world student-AI interactions in programming education. We present a multi-method study of university-level programming courses that combines analysis of 10,632 student–AI dialogue logs collected over three years with student surveys and educator interviews. This setup lets us observe how students naturally use AI assistance (without externally imposed constraints), how those interaction patterns align with metacognitive phases (planning, monitoring, evaluation), and what design principles derive for learner-centered AI scaffoldings that foster metacognitive skill development.

Guided by this motivation, our study addresses the following research questions:
\begin{itemize}

 \item \textbf{RQ1 - Metacognitive Interaction Patterns:} What metacognitive engagement patterns (e.g., planning, monitoring, evaluation) do students exhibit when they interact with AI-powered assistants during programming tasks, and what response types do the assistants provide?

 \item \textbf{RQ2 - Student Perceptions:} How do students perceive AI support for programming learning, including perceived benefits, concerns, and suggestions for improvement?

 \item \textbf{RQ3 - Educator’s Views:} How do educators view students’ metacognitive prompts and the pedagogical potential (and risks) of AI-powered scaffolding in programming courses?

 \item \textbf{RQ4 - Design Principles:} What design principles for pedagogical AI assistants follow from the identified interaction patterns and stakeholder perceptions?
\end{itemize}

Synthesizing answers to our research questions, this paper offers a critical account of the design space for pedagogical AI assistants in programming education. Our analysis illuminates how students interact with AI-powered tools, revealing connections to metacognitive practices. From these insights, we derive actionable design principles for pedagogical AI systems that move beyond solution provision toward fostering planning, monitoring, and reflective evaluation rather than enabling metacognitive offloading.

\section{Related Work}

\subsection{Capabilities of LLMs in Programming Education}

LLMs are reshaping programming education by lowering barriers to help-seeking and automating a broad range of programming tasks, including code generation, debugging, code explanation, and the delivery of personalized feedback \cite{rajala2023call, denny2023conversing}. 

Early research in this area has primarily focused on evaluating the capabilities of LLMs on programming tasks \cite{finnie2023my, phung2023generative, rajala2023call, sarsa2022automatic, savelka2023thrilled}, showing that they are able to address common programming challenges with impressive results. For instance, evaluations of OpenAI Codex and GPT models showed that they could solve many standard CS1 and CS2 exam problems, even outperforming median student performance \cite{finnie-ansley2022robots}. Extending this line of work, Savelka et al. \cite{savelka2023thrilled} evaluated GPT-3 and GPT-4 on 599 programming exercises from three Python courses and found that while GPT-3 failed most assessments, GPT-4 was able to pass entire courses without human assistance. Similarly, Phung et al. \cite{phung2023generative} compared GPT models with human tutors and found that they achieved near human-level performance on Python programming tasks as well as in repairing real-world buggy programs. Other studies highlights LLMs' ability to generate not only functional code but also clear, structured explanations that improve learners’ comprehension. For instance, MacNeil et al. \cite{macneil2023experiences} found that GPT-3’s explanations helped students reason through problems more effectively. Similarly, Leinonen et al. \cite{leinonen2023comparing} showed that LLM-generated explanations were often clearer and more pedagogically useful than student-written ones, though they occasionally suffered from inaccuracy or overconfidence. In the context of feedback, LLMs have shown promise in supporting students by commenting on programming assignments and facilitating the application of theoretical knowledge in practice. Prior work demonstrated that students generally evaluated personalized AI-generated feedback positively \cite{pankiewicz2023large}.  Zhang et al. \cite{zhang2024students} reported that learners in a CS1 Java course generally perceived AI-generated feedback as aligned with the established principles of formative feedback.

\subsection{Student and Educator Perspectives with LLMs}

While early research on LLMs in programming education has primarily emphasized model capabilities, a growing body of work has shifted attention toward the human side. Recent studies have begun to systematically examine how students and educators perceive and experience the use of LLMs in programming education \cite{biswas2023role,humble2023cheaters,yilmaz2023augmented,shoufan2023exploring,ma2024enhancing,ma2024exploring}. This is critical because the effectiveness of LLMs as educational tools depends not only on what they can do, but also on how learners and teachers engage with them and adapt their practices accordingly.

Studies consistently show that students view LLMs as valuable supports in programming education \cite{yilmaz2023augmented,shoufan2023exploring,ma2024enhancing}. They appreciate the immediacy and accessibility of LLMs, emphasizing how quickly it can provide solutions or guidance compared to searching online resources or waiting for human assistance \cite{biswas2023role,yilmaz2023augmented}. Beyond efficiency, students also highlight the clarity of LLM-generated feedback, which helps them understand concepts and debug code more effectively \cite{ma2024enhancing}. At the same time, students remain cautious: many acknowledge that answers can be incorrect or misleading \cite{shoufan2023exploring,ma2024enhancing}. They also worried about the unintended consequences of unmoderated LLM use lead to superficial understanding and lower knowledge retention \cite{skjuve2023user, tlili2023if}. In the absence of deliberate practice, these behaviors risk undermining the development of problem-solving persistence, debugging proficiency, and critical thinking \cite{kasneci2023chatgpt, pankiewicz2023large, shoufan2023exploring}. 

Educators’ perspectives on LLMs have also been widely examined in prior studies. For instructors, such tools can reduce repetitive support work and free time for higher-level teaching, while also providing a private channel for students reluctant to ask questions in public \cite{gao2022who}. Although LLMs are seen as offering unprecedented scalability and personalization, educators tend to approach them with caution \cite{kasneci2023chatgpt, denny2023computing}. Compared to students, educators are generally more skeptical, voicing stronger concerns about over-reliance, misuse, and ethical implications, and calling for institutional policies and guidelines \cite{chan2023ai}. Empirical studies further indicate that teaching practices must adapt as LLMs enter classrooms \cite{amani2023generative}, and highlight the need for educators to remain abreast of technological developments while guiding students on ethical use \cite{prather2023robots}. Some advocate restricting or banning LLM use to preserve academic integrity \cite{lau2023ban}, whereas others experiment with integration strategies that leverage AI for formative support while safeguarding assessment integrity \cite{becker2023programming}. Across these perspectives, there is broad agreement on the importance of pedagogical scaffoldings that encourage strategic help-seeking and foster metacognitive engagement \cite{denny2023computing, sun2024would,kazemitabaar2024codeaid}.

\subsection{LLM-based Coding Assistants}

Concerns about student over-reliance on LLMs, together with evidence that many students struggle to write effective prompts and thus receive unhelpful feedback \cite{ma2024exploring,li2025coderunner}, have led researchers to create custom tools that balance LLM capabilities with pedagogical goals. These tools often incorporate mechanisms such as stepwise hints \cite{kazemitabaar2023studying},  prompt construction for code generation \cite{denny2023promptly}, or restrictions on direct answers \cite{liffiton2023codehelp,kazemitabaar2024codeaid}, ensuring that AI assistance supports learning rather than bypassing it.

For example, Kazemitabaar et al. \cite{kazemitabaar2023studying} created Coding Steps, a tool where students give prompts and receive code automatically generated by LLMs. Yet, they observed that many students simply copied the exercise text as prompts and accepted the AI output without modification. In response to such tendencies, Lifton et al. \cite{liffiton2023codehelp} developed CodeHelp, which supports programming learners by offering scaffolded guidance rather than full solutions. Students can submit questions alongside their code and optional error messages, and the system provides structured hints instead of direct answers. Building further on this idea, CodeAid \cite{kazemitabaar2024codeaid} incorporates pre-defined templates, enabling features such as conceptual explanations, pseudo-code generation with step-by-step commentary, and annotated feedback on incorrect code. 

In sum, current research has made strong progress in building guardrails that keep AI from generating answers directly. However, most studies stop at measuring performance or usability, leaving open questions about how these tools actually shape students’ high-order thinking. In particular, there is little evidence on whether current designs truly support metacognitive development including planning, monitoring, and evaluation. This gap points to the need for research that examines how student interactions with AI align with and support metacognitive processes.

\subsection{Metacognition and Self-regulated Learning in Programming Learning}

Metacognitive skills have long been recognized as essential ability for programming learners \cite{loksa2022metacognition}, where novices often struggle with structuring approaches (planning) \cite{ebrahimi2006taxonomy,hao2025towards}, identifying and fixing errors (monitoring) \cite{park2025exploring}, and optimizing solutions (evaluation) \cite{saliba2024learning}. Prior research has consistently demonstrated the benefits of metacognitive support for learning engagement and outcomes. For example, Choi et al. \cite{choi2023benefit} found that prompting reflection after programming tasks improved performance in both immediate and delayed tests. Similarly, Yilmaz and Yilmaz \cite{karaoglan2022learning} reported higher engagement when students received personalized metacognitive feedback, and Cheng et al. \cite{cheng2024exploring} showed that high-performing students relied on elaboration and critical thinking, while lower-performing peers used more basic strategies.

In the context of human–AI interaction, however, concerns have emerged that easy access to AI-generated solutions may reduce students’ engagement in metacognitive practices, a phenomenon described as “metacognitive laziness” \cite{fan2025beware}. Studies suggest that students sometimes bypass essential metacognitive processes such as evaluation when using AI-powered tools \cite{chen2025unpacking}. While scaffolds such as step-by-step pseudo-code can encourage reflection \cite{kazemitabaar2024codeaid}, overly complete solutions may discourage persistence and exploration \cite{kazemitabaar2023studying}. This tension highlights a critical gap: although the benefits of the metacognition for programming learning are well established, far less is known about how the metacognition phases and strategies evolve when learners interact with AI-powered assistants \cite{viberg2025chatting}. Since metacognition is a dynamic process, strategies may shift in response to AI-mediated interaction. Understanding these dynamics is essential for designing AI-powered pedagogical assistants that reinforce rather than replace students’ metacognitive engagement \cite{phung2025plan}. This gap motivates our study, which analyzes student–AI dialogues to uncover key metacognitive strategies and derive actionable design principles for pedagogical AI systems.


\section{Method}

We conducted a multi-year study in an introductory undergraduate Python programming course at a national university in Japan. The study spanned from 2023 to 2025 and consisted of four offerings of the same course (one offering in 2023, two in 2024, and one in 2025), with a total of 248 students. Throughout each course, all students were given optional access to an AI tool as a supplementary learning resource. Access was provided through a custom, web-based interface that enabled real-time interaction with GPT-based AI. All conversations between students and AI were systematically logged for subsequent analysis. The study adopted a mixed-methods approach, combining quantitative and qualitative data collection. The study received approval from the university’s ethics review board prior to the start of the study.

\subsection{Course Structure}

This first-year undergraduate course, designed for beginners, covered the foundations of the Python programming language in 14 lessons delivered over one semester. Each 90-minute lesson was divided into two segments: the first 45 minutes were dedicated to direct instruction, supported by lecture presentations and slides, followed by 45 minutes of hands-on programming tasks related to the lesson’s topic (e.g., write a specific function based on the requirements). This structure was intended to enable students to immediately apply the theoretical concepts introduced in class. 

Two instructors delivered the four course offerings, and the curriculum was largely unchanged between 2023 and 2025, which allowed us to maintain consistency across offerings. In each lesson, students completed programming exercises. Each course included a total of 57 exercises (averaging 4.75 per lesson). All exercises were submitted through the university’s Learning Management System (LMS), which automatically logged and evaluated each exercise submission, generating a complete and traceable record of students’ programming activity. Students had access to a variety of learning resources beyond AI, including lecture slides and support from the course instructors, and were instructed to cite any external sources used in their work.

\subsection{Participants}

The study involved a total of 248 undergraduate students over a three-year period (62 participants in 2023, 126 in 2024, and 60 in 2025), across four offerings of the same course in the university. Participation was voluntary, informed consent was obtained from all participants, and no aspect of participation affected students’ course grades. The distribution of demographic characteristics was generally consistent across years. Overall, the majority of participants (96.8\%) were first-year students at the time of participation, with a near-equal gender distribution (53.2\% male, 45.8\% female, and 1\% reporting a gender identity outside the male–female binary). While most participants were domestic students from Japan, approximately 14.5\% were international students from diverse countries.

\subsection{Data Sources}

To gather a comprehensive understanding of student interactions with AI and student and educator perceptions of AI, we employed a multifaceted data collection approach by collecting student-AI interaction logs, pre-questionnaire and post-questionnaire from students.


A primary data source for understanding students' usage patterns and AI's response was student-AI dialogue logs. Dialogue data has become an important data source to understand programming experiences \cite{brandt2009two} and coding approaches \cite{finnie2023my, ichinco2015exploring}, particularly when interacting with AI \cite{kazemitabaar2023studying, kazemitabaar2023how_novices_use_llm_code}. For each question asked by students, we closely examined its content and AI generated responses through a thematic analysis. 

%


At the beginning of the course, we asked all participants to fill out a pre-questionnaire to collect demographics such as major and gender. We also asked about their programming experiences and their familiarity with GenAI. As the course progressed, the students' perceptions of using AI in programming education became more evident. To capture these insights, we conducted a questionnaire after the course, gathering students' opinions to understand the advantages and challenges of integrating AI into the programming curriculum from their perspective. We also asked open questions about students’ viewpoints on the use of AI for programming learning purposes, how useful they found it, what they liked or disliked, and any open-ended feedback about it. All the questionnaire items are provided in Appendix \ref{appendix_student}.

\begin{figure}[t]
\centering
\includegraphics[scale=0.45]{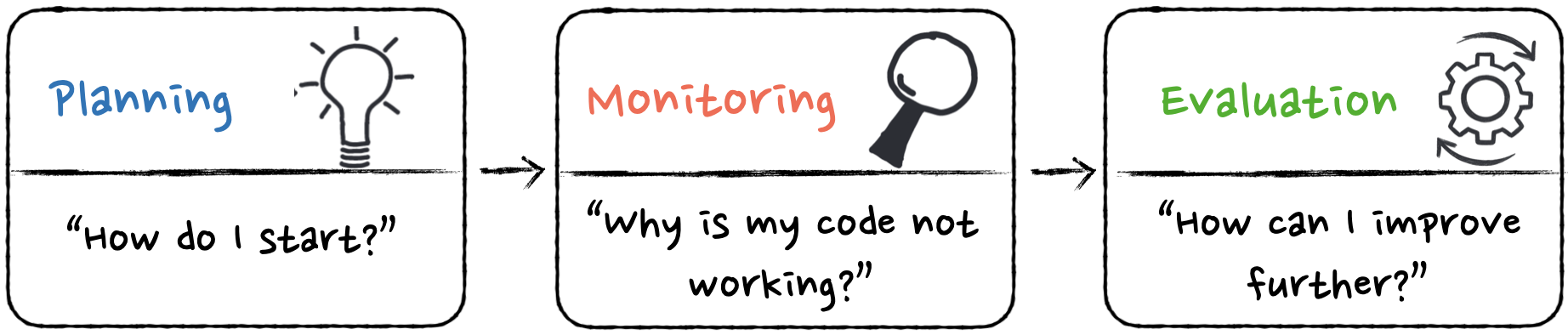}
\caption{Metacognitive Phases of Student–AI Interaction in Programming Tasks}
\label{phases}
\Description{Metacognitive phases in AI-assisted programming. The cycle moves from Planning (goal setting and task decomposition; e.g., “How do I start?”) to Monitoring (debugging and verification; “Why is my code not working?”) and Evaluation (refinement and generalization; “How can I improve further?”). Arrows indicate a progression that can iterate across turns.}
\end{figure}

\begin{table}[t]
\centering
\caption{The sub-dimensions from our thematic analysis, and inter-rater reliability metrics using Cohen's Kappa and percentage agreement. The detailed codebook is provided in Appendix \ref{appendix_codebook}.}
\begin{tabular}{p{0.3\linewidth} >{\raggedright\arraybackslash}p{0.47\linewidth} p{0.15\linewidth}}
    \toprule
    \textbf{Sub-Dimensions} & \textbf{Codes} & \textbf{Inter-Rater Reliability}\\
    \midrule
    What do students ask AI in \textit{Planning} phase? &
    \textit{Conceptual Question (CQ)}, \textit{Problem Understanding (PU)}, \textit{Asking for Example (AE)}, \textit{Code Implementation Questions (CI)}, \textit{Code Generation (CG)}  & 88\% ($\kappa = .79$) \\
      What do students ask AI in \textit{Monitoring} phase?&
    \textit{Error Message Interpretation (EM)}, \textit{Code Correction (CC)}, \textit{Code Verification (CV)}   & 90\% ($\kappa = .85$)  \\
     What do students ask AI in \textit{Evaluation} phase? &
    \textit{Code Explanation (CE)}, \textit{Code Optimization (CO)} &
   100\% ($\kappa = 1.0$)   \\
    \midrule
    How much is AI directly revealing the solution? &
    \textit{Exact Solution Code}, \textit{Steps to Fix Semantic Issue}, \textit{Steps to Fix Syntax Issue}, \textit{Step to Fix External Issue},
    \textit{Example Code}, \textit{Conceptual Explanation}, \textit{Code Explanation}
    & 91\% ($\kappa = .87$) \\
    How technically correct is the response? &
    \textit{Correct}, \textit{Incorrect}
    & 99\% ($\kappa = .85$) \\
    How helpful is the response if correct? &
    \textit{Helpful}, \textit{Not helpful}
    & 100\% ($\kappa = 1$) \\
    \bottomrule
\end{tabular}
\label{highlevel_codebook}
\end{table}

\subsection{Thematic Analysis}

Our analysis sought to capture the student-AI interactions reflect or support students’ metacognitive processes (e.g., planning, monitoring, evaluating). To this end, we conducted a thematic analysis of the dialogue logs. From the 10,632 recorded interactions, we performed stratified random sampling across cohorts, drawing 2,782 instances (26\%) in total. Each year contributed a proportional segment of the sample, ensuring balanced representation for subsequent thematic analysis.

Drawing on coding schemes from prior studies \cite{kazemitabaar2024codeaid,ma2024exploring}, we organized our analysis around two high-level dimensions: student prompts and AI-generated responses (see Table \ref{highlevel_codebook}). This two-part structure enabled us to systematically capture both the student-facing and system-facing aspects of the interaction.

For the first dimension, our aim was to characterize not only the types of questions students asked but also the metacognitive phases embedded in these questions. While earlier codebooks had provided useful structures for describing metacognitive processes in programming contexts \cite{prasad2024self,silva2024learning}, they had not been applied to the setting of student-AI interactions. To address this gap, we extended these prior frameworks to design a codebook that explicitly mapped prompts onto planning, monitoring, and evaluation activities, alongside related help-seeking categories. As shown in Figure \ref{phases}, in the planning phase, students seek support in designing initial solution strategies, which includes help with understanding the problem requirements, exploring potential approaches, and generating starter code or conceptual examples that guide their entry into the task. During monitoring, students seek support in identifying and resolving issues during code execution, which involves assistance with interpreting error messages, verifying program correctness, and debugging both syntactic and semantic problems as they occur. In the evaluation phase, students seek support in reflecting on and improving code quality, which entails guidance on code explanation, optimization, and refinement, enabling learners to go beyond minimal functionality and enhance readability, efficiency, and overall robustness. This codebook allowed us to examine how students’ interactions with AI reflected underlying metacognitive engagement.

The second dimension examined AI-generated responses, emphasizing pedagogical quality and relevance. Building on categories adapted from prior work \cite{kazemitabaar2024codeaid}, we refined and consolidated a codebook and coded each reply on three independent axes: (1) Solution Revelation (ranging from exact solution code; steps to fix semantic, syntax, or external issues; to example code, conceptual explanation, and code explanations), (2) Technical Correctness (correct vs. incorrect), and (3) Helpfulness (helpful vs. not helpful). A key refinement was to treat out-of-scope solutions as Not Helpful even when technically correct—i.e., answers that recommend approaches beyond course expectations or constraints. These clarifications tightened category boundaries and aligned the scheme with our study context.

The coding procedure was identical across both dimensions. Following an inductive approach \cite{bingham2021deductive}, two researchers jointly reviewed an initial set of 800 randomly sampled interactions, generating preliminary sub-dimensions and codes. They then independently coded an additional 150 samples using this preliminary codebook. Then, the results were discussed with the course instructors, and disagreements were resolved through consensus, leading to a refined version of the codebook. To establish reliability, the two researchers independently coded 334 further samples, and inter-rater agreement was calculated using Cohen’s Kappa and percentage agreement \cite{miles1994qualitative,neuendorf2017content}. Once reliability thresholds were satisfied, the researchers proceeded to independently code an additional 2,448 interactions randomly drawn from the remaining dataset. The finalized codebook is provided in Appendix \ref{appendix_codebook}.

\begin{figure}[t]
\centering
\includegraphics[scale=0.45]{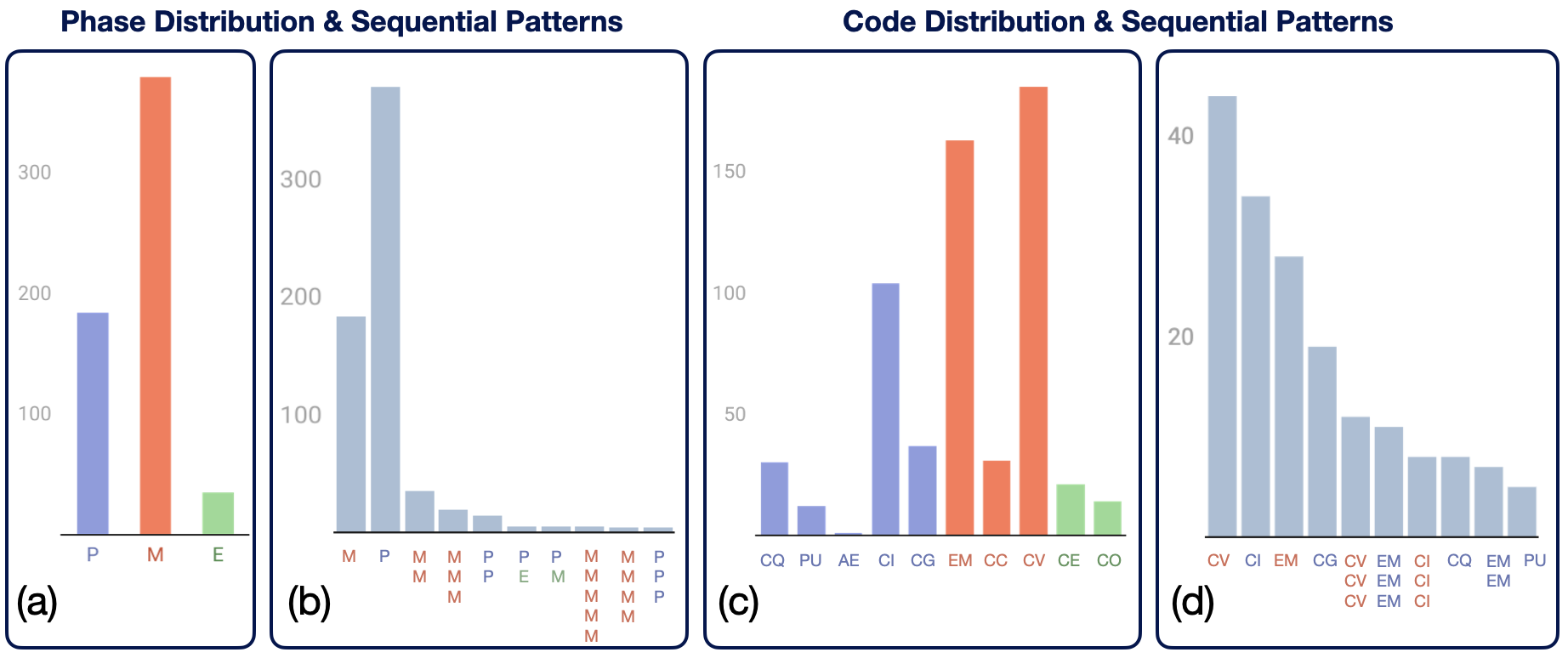}
\caption{Distribution and sequential patterns of student prompts across metacognitive phases and specific categories: (a) distribution of phases, (b) distribution of strategies, (c) sequential phase patterns, and (d) sequential strategy patterns.}
\Description{Phase and strategy distributions and their sequential patterns. (a) Prompts were concentrated in Monitoring (M), far exceeding Planning (P) and Evaluation (E). (b) Within categories, debugging-related prompts such as error message interpretation (EM), code correction (CC), and code verification (CV) dominated. (c) Sequential phase patterns highlight frequent Monitoring-Monitoring loops. (d) Sequential strategy patterns show that verification and debugging tasks appeared repeatedly in succession.}
\label{interaction-1}
\end{figure}

\section{Analysis of Interactions (RQ1)}

\begin{figure}[t]
\centering
\includegraphics[scale=0.43]{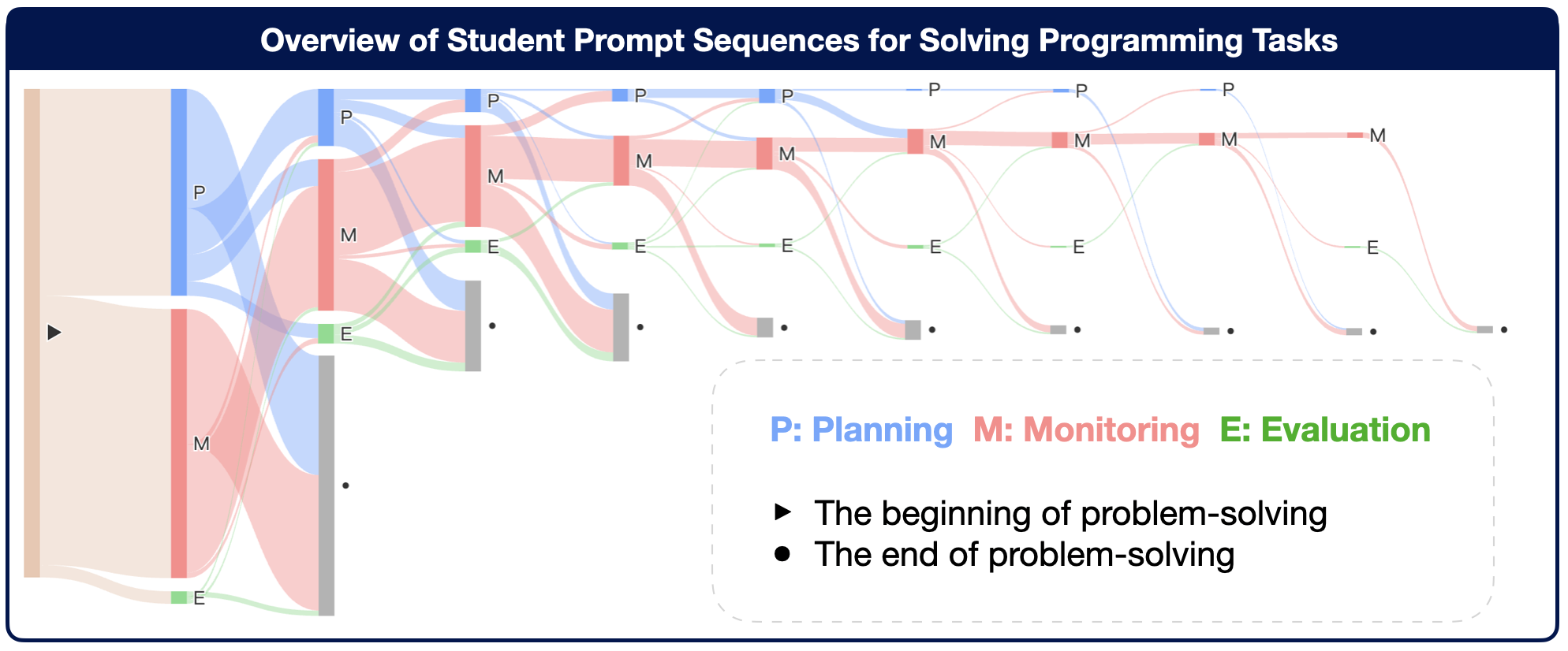}
\caption{Sequences of prompt during problem solving. Each path shows a per-problem trajectory from Start to End. Columns are successive turns. Nodes are grouped as P (Planning), M (Monitoring), and E (Evaluation).}
\Description{The diagram summarizes how students move across metacognitive phases while solving programming tasks with AI. Sequences typically begin and then persist in Monitoring, with fewer transitions back to Planning or forward to Evaluation; only a subset of trajectories reach Evaluation before ending.}
\label{interaction-2}
\end{figure}

\subsection{Analysis of Student Prompts}

To understand how students engaged with AI across different stages of problem solving, we first examined the distribution of prompts by metacognitive phase and strategies. Figure~\ref{interaction-1}(a) presents the overall distribution of student prompts across the three metacognitive phases. Monitoring (M) accounted for the largest share of prompts, far exceeding Planning (P) and Evaluation (E). This indicates that students most frequently sought AI assistance during active debugging, troubleshooting, and correctness verification, rather than during initial strategy formulation or reflective evaluation. As shown in Figure~\ref{interaction-1}(b), within the strategies, Error Message Interpretation (EM), Code Correction (CC), and Code Verification (CV) dominated, highlighting students’ tendency to rely on AI as a just-in-time debugging tool. In contrast, prompts associated with planning activities, such as Conceptual Questions (CQ), Problem Understanding (PU), or Asking for Example (AE), were comparatively sparse, and evaluation-related prompts (Code Explanation and Code Optimization) were least frequent. These descriptive patterns suggest that students primarily positioned AI as a reactive aid to overcome immediate coding obstacles, rather than as a proactive partner for strategic planning or reflective learning.

To examine how prompts unfolded over time, Figure~\ref{interaction-2} visualizes sequential flows with a Sankey diagram. A substantial share of sessions began directly in Monitoring rather than Planning, indicating that students often approached AI reactively after encountering execution problems rather than proactively during problem framing. Evaluation prompts often appeared as terminal moves, occurring at the end of problem-solving sequences, consistent with the notion that reflection was an afterthought rather than an integrated process. These patterns demonstrate that student–AI interactions tended to be short-sighted and narrowly focused on immediate code repair. 

Figure~\ref{interaction-1} (c–d) further illustrates sequential patterns of student prompts in phases and strategies. At the phase level (Figure~\ref{interaction-1}(c)), most frequent prompt sequences consist of a single phase, with monitoring are the most, followed by planning, while evaluation were rare. The majority of sequences were short and dominated by repetitive Monitoring, confirming the persistence of debugging-focused interactions. Longer or cross-phase sequences were comparatively rare, with only a few Planning-Monitoring-Evaluation trajectories, indicating that students seldom cycled through a full metacognitive loop with AI. At the strategy level (Figure~\ref{interaction-1}(d)), the most frequent transitions involved Code Verification (CV), Code Implementation Question (CI), and Error Message Interpretation (EM), often chained in quick succession. These transitions suggest that students tended to oscillate between checking correctness, requesting implementation, and interpreting errors, without extending into broader conceptual or evaluative strategies. Collectively, the sequential analyses reinforce the picture of students treating AI primarily as an immediate error-fixing assistant, with limited movement toward higher-order reflection or iterative planning improvement.

\begin{figure}[tb]
\centering
\includegraphics[scale=0.4]{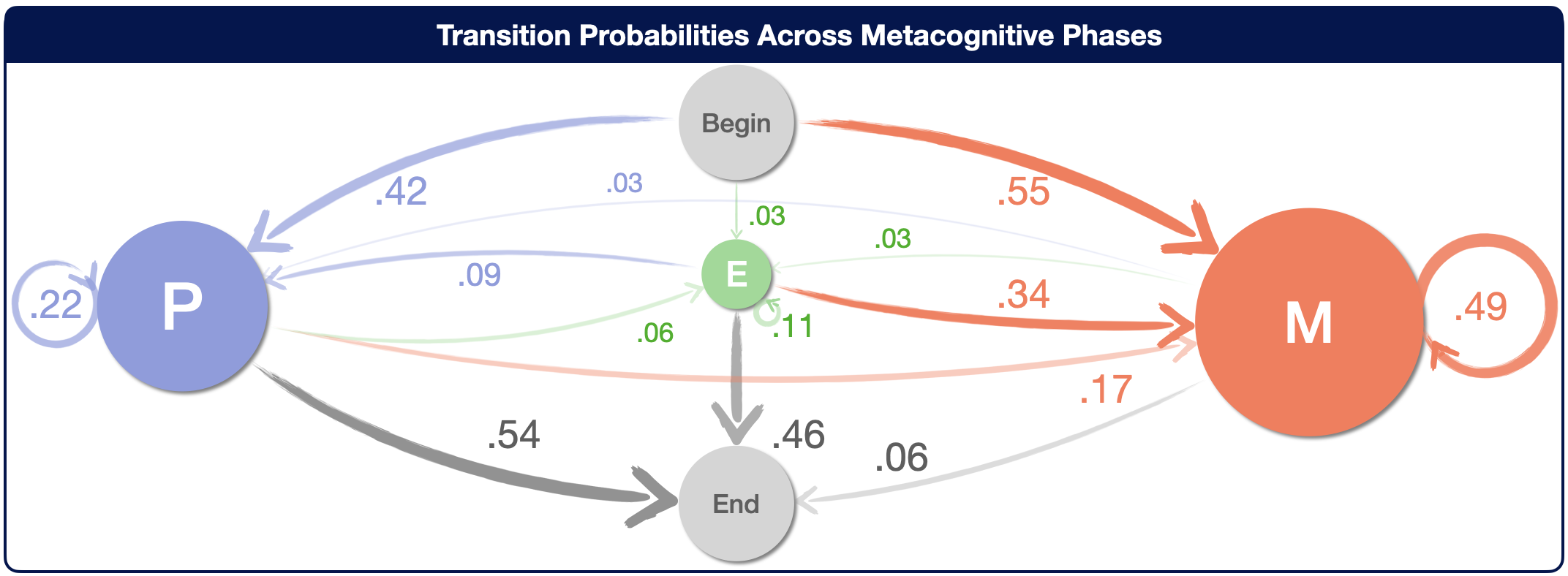}
\caption{Markov Chain Analysis of Phase Transitions.}
\label{interaction-3}
\Description{Transition probabilities across Planning (P), Monitoring (M), and Evaluation (E). Monitoring exhibited the strongest self-loop (0.49), indicating students’ tendency to remain in debugging-focused interactions. The dominant cross-phase pathway was Planning-Monitoring (0.54), reflecting a quick shift from initial framing to troubleshooting. By contrast, transitions into Evaluation were weak (Monitoring-Evaluation: 0.17; Planning-Evaluation: 0.06), and nearly half of the sessions terminated after Monitoring (0.46), underscoring the marginal role of reflective evaluation in students’ workflows.}
\end{figure}

Finally, Figure~\ref{interaction-3} quantifies these patterns using a Markov chain model. Monitoring showed the highest self-transition probability (0.49), confirming its stickiness. Planning also exhibited a self-loop (0.22), though weaker, suggesting some iteration in problem framing. Most cross-phase transitions converged on Monitoring, the dominant cross-phase pathway was Evaluation-to-Monitoring (0.54) and Planning-to-Monitoring (0.17). By contrast, transitions into Planning and Evaluation were weak, and nearly half of the sessions terminated after Planning (0.34) and Evaluation (0.46). Together, these results underscore that students positioned AI primarily as a reactive debugging assistant, with Planning under-specified and Evaluation marginal, leaving metacognitive phases fragmented rather than integrated.

\subsection{AI-generated Responses}

\subsubsection{Response Types across Phases and Strategies} 

Table~\ref{phase_category_response} details the types of responses the AI tended to produce across phases. In Planning, outputs were dominated by Example Code (S5, 48.3\%) and Exact Solution Code (S1, 35.2\%), with smaller proportions of Conceptual Explanation (S6, 14.2\%) and very little Code Explanation (S7, 2.3\%). This pattern aligns with the strategy-level distribution: most Planning prompts elicited examples or conceptual scaffolds, yet Code Generation (CG) consistently yielded runnable code, while Code Implementation (CI) was split—some responses provided direct solutions, but a larger share consisted of illustrative examples.

Monitoring exhibited a markedly different profile in which the model tended to “patch first.” Here, Exact Solution Code (S1) accounted for 51.7\% of responses and Steps to Fix Syntax Issues (S3) for 30.3\%. Within Monitoring categories, Error Message Interpretation (EM) most often led to syntax-fix steps, whereas both Code Correction (CC) and Code Verification (CV) were dominated by direct runnable code.

In Evaluation, responses shifted toward interpretive support. The phase showed the highest share of Code Explanation (S7, 38.7\%), alongside Exact Solution Code (S1, 29.0\%), Example Code (S5, 16.1\%), and Conceptual Explanation (S6, 12.9\%). At the strategy-level, Code Explanation (CE) was almost entirely composed of Code Explanation (S7), while Code Optimization (CO) predominantly produced Exact Solution Code (S1).

\begin{table}[t]
\small
\centering
\caption{Distribution of response types (S1--7) across phases and strategies.}
\label{phase_category_response}
\begin{tabular}{llrrrrrrr}
\toprule
Phase & Category & S1 & S2 & S3 & S4 & S5 & S6 & S7 \\
\midrule
\multirow{6}{*}{Planning (P)}

& Conceptual Question (CQ) & \cellcolor[RGB]{248,252,254}{6.9\%} & \cellcolor[RGB]{255,255,255}{0.0\%} & \cellcolor[RGB]{255,255,255}{0.0\%} & \cellcolor[RGB]{255,255,255}{0.0\%} & \cellcolor[RGB]{190,221,253}{34.5\%} & \cellcolor[RGB]{121,183,252}{51.7\%} & \cellcolor[RGB]{248,252,254}{6.9\%} \\

& Problem Understanding (PU) & \cellcolor[RGB]{255,255,255}{0.0\%} & \cellcolor[RGB]{255,255,255}{0.0\%} & \cellcolor[RGB]{255,255,255}{0.0\%} & \cellcolor[RGB]{255,255,255}{0.0\%} & \cellcolor[RGB]{194,223,253}{33.3\%} & \cellcolor[RGB]{124,185,252}{50.0\%} & \cellcolor[RGB]{235,244,253}{16.7\%} \\

& Asking for Example (AE) & \cellcolor[RGB]{255,255,255}{0.0\%} & \cellcolor[RGB]{255,255,255}{0.0\%} & \cellcolor[RGB]{255,255,255}{0.0\%} & \cellcolor[RGB]{255,255,255}{0.0\%} & \cellcolor[RGB]{30,144,255}{100.0\%} & \cellcolor[RGB]{255,255,255}{0.0\%} & \cellcolor[RGB]{255,255,255}{0.0\%} \\

& Code Implementation Question (CI) & \cellcolor[RGB]{222,237,252}{23.7\%} & \cellcolor[RGB]{255,255,255}{0.0\%} & \cellcolor[RGB]{255,255,255}{0.0\%} & \cellcolor[RGB]{255,255,255}{0.0\%} & \cellcolor[RGB]{88,168,251}{72.2\%} & \cellcolor[RGB]{250,252,254}{4.1\%} & \cellcolor[RGB]{255,255,255}{0.0\%} \\

& Code Generation (CG) & \cellcolor[RGB]{30,144,255}{100.0\%} & \cellcolor[RGB]{255,255,255}{0.0\%} & \cellcolor[RGB]{255,255,255}{0.0\%} & \cellcolor[RGB]{255,255,255}{0.0\%} & \cellcolor[RGB]{255,255,255}{0.0\%} & \cellcolor[RGB]{255,255,255}{0.0\%} & \cellcolor[RGB]{255,255,255}{0.0\%} \\

& \textbf{Overall} & \cellcolor[RGB]{187,220,253}{\textbf{35.2\%}} & \cellcolor[RGB]{255,255,255}{\textbf{0.0\%}} & \cellcolor[RGB]{255,255,255}{\textbf{0.0\%}} & \cellcolor[RGB]{255,255,255}{\textbf{0.0\%}} & \cellcolor[RGB]{130,189,252}{\textbf{48.3\%}} & \cellcolor[RGB]{239,246,253}{\textbf{14.2\%}} & \cellcolor[RGB]{254,255,255}{\textbf{2.3\%}} \\
\midrule
\multirow{4}{*}{Monitoring (M)}

& Error Message Interpretation (EM) & \cellcolor[RGB]{213,233,253}{30.1\%} & \cellcolor[RGB]{252,254,255}{5.1\%} & \cellcolor[RGB]{120,182,252}{51.3\%} & \cellcolor[RGB]{255,255,255}{1.9\%} & \cellcolor[RGB]{244,249,253}{10.9\%} & \cellcolor[RGB]{255,255,255}{0.6\%} & \cellcolor[RGB]{255,255,255}{0.0\%} \\

& Code Correction (CC) & \cellcolor[RGB]{62,160,252}{80.0\%} & \cellcolor[RGB]{254,255,255}{3.3\%} & \cellcolor[RGB]{243,249,253}{10.0\%} & \cellcolor[RGB]{255,255,255}{0.0\%} & \cellcolor[RGB]{255,255,255}{0.0\%} & \cellcolor[RGB]{255,255,255}{0.0\%} & \cellcolor[RGB]{248,252,254}{6.7\%} \\

& Code Verification (CV) & \cellcolor[RGB]{82,170,251}{66.5\%} & \cellcolor[RGB]{255,255,255}{0.0\%} & \cellcolor[RGB]{238,246,253}{14.7\%} & \cellcolor[RGB]{255,255,255}{0.0\%} & \cellcolor[RGB]{255,255,255}{0.0\%} & \cellcolor[RGB]{255,255,255}{1.2\%} & \cellcolor[RGB]{240,247,253}{17.6\%} \\

& \textbf{Overall} & \cellcolor[RGB]{118,181,252}{\textbf{51.7\%}} & \cellcolor[RGB]{254,255,255}{\textbf{2.5\%}} & \cellcolor[RGB]{215,234,253}{\textbf{30.3\%}} & \cellcolor[RGB]{255,255,255}{\textbf{0.8\%}} & \cellcolor[RGB]{252,254,255}{\textbf{4.8\%}} & \cellcolor[RGB]{255,255,255}{\textbf{0.8\%}} & \cellcolor[RGB]{241,248,253}{\textbf{9.0\%}} \\
\midrule
\multirow{3}{*}{Evaluation (E)}
& Code Explanation (CE) & \cellcolor[RGB]{255,255,255}{0.0\%} & \cellcolor[RGB]{255,255,255}{0.0\%} & \cellcolor[RGB]{253,254,255}{5.3\%} & \cellcolor[RGB]{255,255,255}{0.0\%} & \cellcolor[RGB]{239,247,253}{15.8\%} & \cellcolor[RGB]{239,247,253}{15.8\%} & \cellcolor[RGB]{94,172,252}{63.2\%} \\

& Code Optimization (CO) & \cellcolor[RGB]{69,163,252}{75.0\%} & \cellcolor[RGB]{255,255,255}{0.0\%} & \cellcolor[RGB]{255,255,255}{0.0\%} & \cellcolor[RGB]{255,255,255}{0.0\%} & \cellcolor[RGB]{239,247,253}{16.7\%} & \cellcolor[RGB]{245,250,254}{8.3\%} & \cellcolor[RGB]{255,255,255}{0.0\%} \\
& \textbf{Overall} & \cellcolor[RGB]{212,232,253}{\textbf{29.0\%}} & \cellcolor[RGB]{255,255,255}{\textbf{0.0\%}} & \cellcolor[RGB]{252,254,255}{\textbf{3.2\%}} & \cellcolor[RGB]{255,255,255}{\textbf{0.0\%}} & \cellcolor[RGB]{241,248,253}{\textbf{16.1\%}} & \cellcolor[RGB]{244,249,253}{\textbf{12.9\%}} & \cellcolor[RGB]{159,202,253}{\textbf{38.7\%}} \\
\bottomrule
\end{tabular}

\bigskip
\footnotesize \textit{Note.} Response types:
S1 = Exact Solution Code,
S2 = Step to Fix Semantic Issue,
S3 = Step to Fix Syntax Issue,
S4 = Step to Fix External Issue,
S5 = Example Code,
S6 = Conceptual Explanation,
S7 = Code Explanation.
\end{table}

\subsubsection{Correctness of AI Responses}

As shown in Table~\ref{correctness_helpfulness}, the overall correctness rate of AI responses was high, but uneven across phases and categories. Planning responses (98.9\%) were almost always correct, likely because conceptual clarifications and small examples leave less room for factual error. Evaluation responses (100\%) similarly achieved perfect correctness, as explanation and commentary tasks required little speculative inference. By contrast, Monitoring responses (94.1\%) displayed notable vulnerabilities, particularly in Code Correction (80\% correct). A recurrent issue was overcorrection: the AI attempted to “improve” code by introducing extra modifications that were unnecessary or even harmful, sometimes converting a correct snippet into a faulty one. This failure mode underscores a fundamental challenge of LLMs in educational contexts—the tendency to produce overconfident edits when the minimal fix would suffice. The result is that while Monitoring responses were often technically competent, their reliability was uneven, making them less trustworthy for novice learners who lack the expertise to discriminate between accurate and overextended corrections.

\begin{table}[t]
\centering
\caption{Correctness and helpfulness rates by phase and category.}
\label{correctness_helpfulness}
\begin{tabular}{llrr}
\toprule
Phase & Category & Correct (\%) & Helpful (\%) \\
\midrule
\multirow{6}{*}{Planning (P)} 
  & Conceptual Question (CQ)      & 100.0\% & 89.7\% \\
  & Problem Understanding (PU)    & 100.0\% & 100.0\% \\
  & Asking for Example (AE)          & 100.0\% & 100.0\% \\
  & Code Implementation Question (CI)      &  97.9\% & 89.7\% \\
  & Code Generation (CG)          & 100.0\% & 100.0\% \\
  & \textbf{Phase Overall}        & \textbf{98.9\%} & \textbf{92.6\%} \\
\midrule
\multirow{4}{*}{Monitoring (M)} 
  & Error Message Interpretation (EM) & 95.5\% & 91.0\% \\
  & Code Correction (CC)              & 80.0\% & 80.0\% \\
  & Code Verification (CV)            & 95.3\% & 89.4\% \\
  & \textbf{Phase Overall}            & \textbf{94.1\%} & \textbf{89.3\%} \\
\midrule
\multirow{3}{*}{Evaluation (E)} 
  & Code Explanation (CE) & 100.0\% & 94.7\% \\
  & Code Optimization (CO)           & 100.0\% & 91.7\% \\
  & \textbf{Phase Overall}           & \textbf{100.0\%} & \textbf{93.5\%} \\
\bottomrule
\end{tabular}
\end{table}

\subsubsection{Helpfulness of AI Responses}

Helpfulness ratings (Table~\ref{correctness_helpfulness}) revealed a second dimension beyond technical correctness: whether responses aligned with students’ learning needs. Planning responses (92.6\%) and Evaluation responses (93.5\%) were rated most helpful, since they provided clarifications, conceptual framing, and interpretive insights that students could directly use to advance their reasoning. Monitoring responses (89.3\%), however, lagged behind. Even when technically correct, many were marked unhelpful when they supplied solutions beyond course scope (e.g., unnecessarily advanced algorithms or libraries) or when they required contextual details that students had not provided (e.g., incomplete I/O specifications). These cases illustrate a crucial distinction: correctness alone does not guarantee pedagogical usefulness. An answer can be “right” yet still fail to move the learner forward in a productive way. Conversely, answers judged unhelpful often reinforced passivity: by handing students runnable code, the AI risked bypassing the reasoning steps that are central to programming education. This reveals a pedagogical tension: effective AI support must balance accuracy with calibration to learner context, encouraging incremental progress without substituting for critical thinking.

\section{Student Perspectives (RQ2)}

\subsection{Analysis of Pre-Questionnaire}

The pre-questionnaire revealed that most participants (90\%) self-identified as beginners in Python programming, and this proportion remained relatively consistent across all three years of the study. Reported usage of AI increased sharply over the three years of our study, reflecting both the rapid advancement of AI technology and its growing integration into students’ everyday learning practices. In 2023, only 36.8\% of respondents reported active use, while 57.9\% indicated non-use. By 2024, active use had risen to 63.9\%, with non-use dropping to 11.1\%. By 2025, this trend became even more pronounced: 91.7\% of respondents reported active use, and only 8.3\% reported non-use. Similarly, students’ self-perceived understanding of AI increased consistently across the study period. In 2023, only 5.2\% of respondents indicated that they could explain AI in detail, 68.4\% reported somewhat understanding, and 26.3\% indicated little or no understanding. By 2024, 47.2\% reported being able to explain AI in detail or to some extent, another 47.2\% reported somewhat understanding, and only 5.6\% indicated low understanding, with no students reporting no knowledge. In 2025, 100\% of respondents indicated at least some understanding, and none reported low or no understanding. Finally, students’ perceptions of the impact of AI on their learning were generally positive and increased across years. Overall, 82.3\% of respondents considered AI to be beneficial or somewhat beneficial, with the proportion rising from 78.9\% in 2023 to 80.5\% in 2024, and 100\% in 2025.

\begin{figure}[t]
\centering
\includegraphics[scale=0.54]{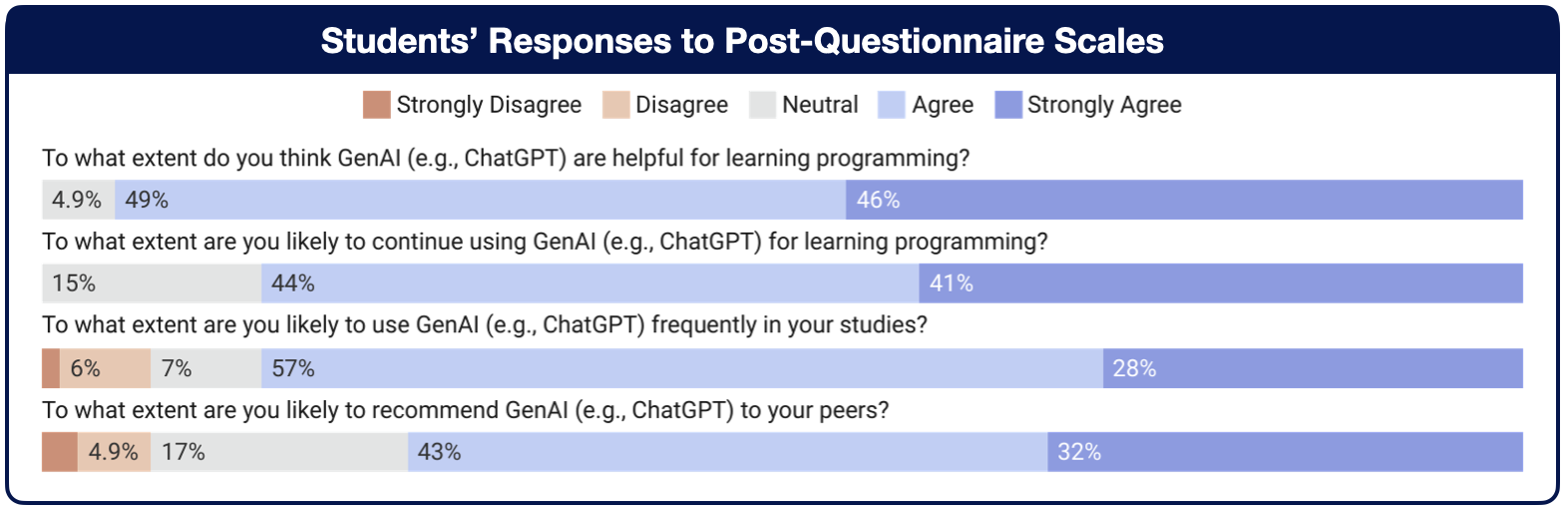}
\caption{Distribution of students’ responses across all post-questionnaire scales.}
\Description{Students’ responses to post-questionnaire Likert scales about GenAI for programming.
Stacked bars show the distribution from Strongly Disagree to Strongly Agree. Most students viewed AI as helpful (49\% Agree, 46\% Strongly Agree, 5\% in non-agreement categories). A large majority intended to continue using AI for learning (44\% Agree, 41\% Strongly Agree, 15\% Neutral). Expected frequency of use was similarly high (57\% Agree, 28\% Strongly Agree, 13\% Disagree/Strongly Disagree). Willingness to recommend GenAI to peers was positive but less unanimous (32\% Strongly Agree, 43\% Agree, 22\% Disagree/Strongly Disagree).}
\label{pre_questionnaire}
\end{figure}

\subsection{Analysis of Post-Questionnaire}

Figure \ref{pre_questionnaire} presents an overview of student responses to the post-questionnaire. The results indicate strong support for AI as helpful tool in learning programming, with 45.7\% completely agreeing and 49.4\% agreeing. Similarly, the intention to continue using ChatGPT for learning programming remains high, with 40.7\% completely agreeing and 44.4\% agreeing. Furthermore, 85.2\% of respondents indicated they would often use AI in the future. Finally, 75.3\% expressed willingness to recommend AI to their friends, highlighting its perceived value beyond personal use. 

Additionally, the results illustrates students’ perceptions of how AI can assist in programming learning, allowing for multiple selections. Providing examples of programming code was the most frequently chosen option (27\%), this was closely followed by answering programming questions (24\%). Explaining programming concepts (19\%) and correcting programming code (17\%) were also commonly reported, while 14\% selected offering learning advice and resources.

 \subsubsection{Benefits of Using AI in Programming Learning}
 
Our analysis of students’ open-ended responses revealed several perceived benefits when using AI tools such as ChatGPT for programming learning.

\paragraph{Personalized and Accessible Support:} Students emphasized the accessibility and personalized nature of AI support. They appreciated being able to ask questions “anytime and anywhere,” often in conversational or vague terms that would be difficult to phrase in a search engine. One respondent noted that “even if I don’t know the right technical words, I can still get a useful answer,” while another highlighted the feeling of “one-to-one support tailored to my needs.”

\paragraph{Error Correction:} One of the most frequently mentioned benefits is the ability of AI to quickly identify and correct errors in students’ code. Rather than spending excessive time debugging on their own, students can receive instant feedback, which both saves time and reduces frustration, allowed them to resolve doubts without waiting for help from instructors or peers. As one respondent described, “when I have a question, it solves it immediately,” while another emphasized that they could “move smoothly to the next task without being stuck.” Such immediacy was often framed as a key factor in keeping up learning momentum.

\paragraph{Concept and Code Explanation:} Another recurring advantage is that AI provide clear explanations of programming concepts and code logic. Many respondents reported that it could explain programming concepts in simpler terms, often accompanied by concrete usage examples. For instance, students described how it “breaks down complex concepts step by step” and “explains why certain functions are used”. Others appreciated how it supplemented their understanding of Python rules they had previously overlooked, “things I couldn’t understand on my own became obvious after the explanation.” Students described situations where they previously struggled to understand why certain code behaved in a particular way, but through explanations, it became much clearer. 

\paragraph{Expanding Ideas and Examples:} In addition, AI support creative thinking by suggesting alternative solutions, offering additional examples, and even suggeting new exercises. This encourages students to explore different approaches rather than relying on a single solution. Several participants reported that AI “not only answered but also proposed alternative methods”, while another reflected, “I thought it would just show me one way to do things, but it pointed out where I was wrong and suggested other methods, which made it easier to understand”. Others noted that it provided not only code but also practical examples of function usage, helping them apply knowledge more flexibly.

\subsubsection{Concerns of Using AI in Programming Learning}

\paragraph{Accuracy and Reliability Issues:} Students repeatedly pointed out that AI responses are not guaranteed to be correct: “You don’t know if that is the correct answer or not because ChatGPT sometimes gives false answers” and “It may not always be accurate and can create mistakes in its responses.” Some noted that the model sometimes “answers based on the question I asked before which has nothing to do with my new questions,” causing confusion. Others mentioned that errors can be subtle and difficult to detect: “If we just prompt the question without thinking first, we wouldn’t learn anything and might absorb wrong information without noticing.”

\paragraph{Overreliance and Reduced Thinking:} Another concern is over-dependence, many worried that “getting the answer too quickly makes me stop thinking for myself” or “I just copy the code without reading it, so I don’t gain any knowledge.” Others echoed this concern with remarks like “ChatGPT is too useful, it makes me lazy,” and “If I rely on ChatGPT all the time, my problem-solving ability will decline.” Several stressed that the danger lies not in occasional use but in the habit of outsourcing every step: “I rely too much and my thinking power decreases” and “If you always ask ChatGPT, you stop trying to solve the problem on your own.”

\paragraph{Lack of Context:} Students also highlighted the model’s inability to account for course-specific constraints. For example, one noted, “it cannot know what the topic we are learning in class, so it might give answers that students have not yet encountered.” Others commented that the answers sometimes diverged from what the instructor expected: “Sometimes the direction from AI is different from the teacher’s.” 
 
 \subsubsection{Suggestions for Improving AI' Support in Programming Learning}
 
\paragraph{Accuracy and Reliability of Responses:} The most consistent demand from students was for more dependable and auditable responses. Participants repeatedly noted that while AI can produce fluent answers, its tendency to “sound confident while being wrong” undermines trust. Typical suggestions included “quantify answer accuracy” and “flag the parts that might be wrong”. Others wanted it to fact-check before responding—“It should check facts before it provides information”, and to continuously update knowledge bases to avoid outdated examples.

In programming contexts, students explicitly asked for verification by execution: “Automatically run generated code in a sandbox and use the results for feedback learning.” Transparency was also emphasized: “Show the reasoning process that led to the answer.” In essence, students are asking not only for accurate output, but for mechanisms of accountability, so that they can gauge whether an answer can be safely trusted.

\paragraph{Instructional Scaffolding and Learning Process Support:} Beyond correctness, students wanted AI to act as a learning companion rather than a shortcut to solutions. Many stressed that the system should resist giving full answers too quickly, instead adopting a scaffolded approach: “Instead of the complete answer, provide hints step by step.” Several requested adaptive questioning to check comprehension—“Ask just-right questions to test if I actually understand.”

This also extended to encouraging productive struggle. For instance: “Don’t show the answer right away; push me to think first.” Some even suggested structural restrictions: “Make students go through the reasoning process before they can see the answer” or “Limit how many times we can ask questions so that we think more on our own.”

Students also wanted multiple solution paths and richer learning aids. Comments included: “Don’t just give one example—show different patterns too” and “Explain with detailed charts or diagrams.” These requests suggest that students are envisioning AI as interactive tutor—not merely delivering answers, but scaffolding reasoning, broadening exposure to alternative approaches to deepen understanding.

\paragraph{Context Awareness, Personalization, and User Experience:} Another cluster of requests focused on context sensitivity and seamless integration into learning environments. Students expressed frustration when AI produced generic or irrelevant answers and asked for alignment with class materials: “I hope it provides code using the methods the teacher taught us.” Others emphasized focus and stability: “It should stay on topic to the question being asked.”

There was also a call for better prompting support and proactive guidance, e.g., “Help me phrase my question so I can get the answer I want” or “Suggest options without me having to ask first.” In terms of tools, integration with existing workflows was highlighted: “Embed it into IDE so errors and fixes appear directly” and “Allow access to my local files.”

Finally, students imagined multimodal and accessible interfaces: “Explain not only with words but also with illustrations,” and “Add voice commands or visual aids.” Together, these comments point toward a vision of AI as personalized, contextually grounded, and embedded in the actual programming environment, rather than a detached Q\&A tool.

\section{Educator Survey (RQ3)}

We surveyed eleven computing educators from Japan and China to understand their perspectives on using GenAI in programming education. The survey began by exploring the educators' backgrounds and their current challenges and strategies, especially around students' utilization of AI-based coding tools. The suvey items are provided in Appendix \ref{appendix_educator}.

All participants were actively involved in teaching computing courses as well as in computer science education research. Their teaching experience varied: two had more than ten years, two had 3–5 years, three had 1–3 years, and four had less than one year. With respect to teaching level, seven participants taught only undergraduate courses, two taught only graduate courses, and two were engaged in both. Regarding course types, three taught only introductory programming, four taught only data science or machine learning, three taught both introductory programming and data science/ML, and one reported teaching all three types (introductory programming, data science/ML, and PBL). As for programming languages, most participants reported teaching with Python, while three also used C/C++, and one reported Java.

\subsection{AI-Policy and General Perceptions}

Educators we surveyed generally reported either “mostly open” or “allowed with conditions” policies toward students’ use of AI. Under conditional policies, students were encouraged to consult AI but were explicitly prohibited from directly generating or submitting complete solutions or code. Some instructors required evidence of prior problem-solving effort before any code generation—“If they want to generate code, they need to show that they have done sufficient work towards solving the problem, e.g., decomposing a problem to sub-problems and using AI to generate code for sub-problems”—and others stressed explicit disclosure/acknowledgment of AI use. By contrast, more open policies emphasized AI’s value for personalized learning and for fostering student responsibility and autonomy, particularly at the graduate level. Across both groups, common restrictions included prohibiting AI use during examinations and discouraging reliance on AI for final answers. Overall, educators expressed generally positive views on the role of AI assistants in student learning: 80\% of the ten respondents either strongly agreed or agreed that AI assistants help students learn programming, while the remaining 20\% were neutral.

\begin{figure}[t]
\centering
\includegraphics[scale=0.37]{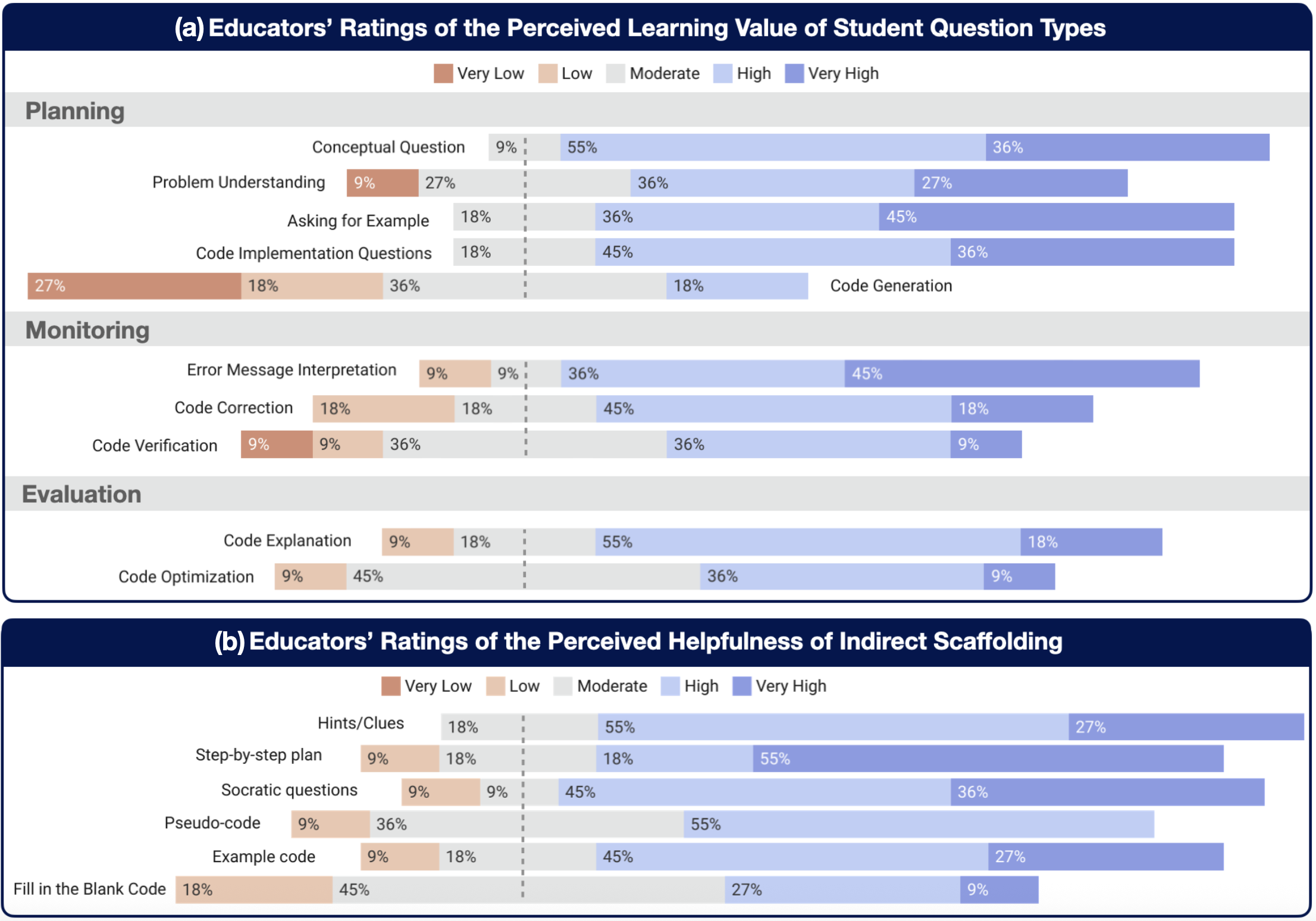}
\caption{Educators’ ratings of (a) different student question types, and (b) the Perceived Helpfulness of Indirect Scaffolding regarding their perceived learning value.}
\label{educator}
\Description{Educators’ ratings of question types and indirect scaffolds.
(a) Perceived learning value of student question types by phase. In Planning, educators rated conceptual questions and asking for examples highly (55\% High/36\% Very High and 36\% High/45\% Very High, respectively), as well as code-implementation questions (45\% High/36\% Very High), while code generation was seen as low value (27\% Very Low, 18\% Low; 18\% High). In Monitoring, error-message interpretation received the strongest endorsement (36\% High/45\% Very High); code correction skewed High (45\%) with some Very High (18\%); code verification was mostly Moderate/High. In Evaluation, code explanation was valued (55\% High/18\% Very High), whereas code optimization was mixed (Large Low 45\% but also High/Very High 45\%).
(b) Perceived helpfulness of indirect scaffolding. Hints/clues and step-by-step plans were rated most helpful. Socratic questions were also strong (45\% High/36\% Very High). Example code was viewed positively (45\% High/27\% Very High), pseudo-code was more mixed (55\% High), and fill-in-the-blank code was least favored (45\% Low; High+Very High 36\%).}
\end{figure}

\subsection{Student Input and Context}

\subsubsection{Perceived Learning Value of Questions}

To further illustrate educators’ perceptions of different types of student prompts, Figure~\ref{educator}(a) summarizes their ratings of the learning value associated with each category across the \textit{planning}, \textit{monitoring}, and \textit{evaluation} phases. Educators consistently associated higher learning value with questions that promote sense-making—such as clarifying concepts, diagnosing errors, and articulating reasoning—than with those that primarily generate complete solutions. Within \textit{planning}, conceptual questions were rated as most beneficial, followed by example requests and implementation questions, and code generation was rated lowest. During \textit{monitoring}, educators valued error interpretation and code correction, whereas code verification elicited more divided views. In \textit{evaluation}, code explanation was recognized as pedagogically useful, and code optimization was often viewed with skepticism, likely reflecting its limited applicability within typical coursework. Collectively, educators believe that AI support and instructional practice should emphasize prompts that foster conceptual framing, diagnostic reasoning, and iterative refinement, rather than encouraging direct solution generation.

\subsubsection{Input Formats for AI Use}

When asked which input format better promotes high-quality thinking when students use AI, most educators favored structured forms designed by instructors or experts, while a smaller number supported free-form inputs. Proponents of structured formats emphasized that such scaffolds can help students articulate their understanding, avoid incomplete or misleading prompts, and improve programming skills. One educator noted that “many students struggle to provide clear, fully contextualized prompts, which often yield misleading outputs. Well-designed, structured input forms (e.g., templates or checklists) can scaffold better prompts and improve comprehension.” Others highlighted that structured formats are particularly important for students with weak programming backgrounds, as they might otherwise resort to directly asking AI for answers without sufficient reasoning. In contrast, educators who preferred free-form inputs pointed to accessibility and student autonomy. They argued that free-form interactions encourage more frequent questioning and that “asking AI effectively is also an important skill.” One educator noted that because free-form input mirrors real-world tools such as ChatGPT, it can lower barriers and promote engagement, even if the questions vary in quality.

\subsection{AI Responses and Scaffolding}

\subsubsection{Educator Preferences for Response Scaffolding}

All educators consistently agreed that AI’s responses should be delivered in the form of indirect scaffolding rather than providing direct runnable code. This shared preference reflects a common belief that scaffolding supports deeper engagement and prevents students from bypassing critical reasoning processes.

Building on this overall preference for indirect scaffolding, Figure~\ref{educator}(b) presents educators’ ratings of different forms of such scaffolding. Hints/clues, along with step-by-step plans and Socratic questioning, were rated most helpful, highlighting the value of guidance that supports reasoning and incremental problem solving. In contrast, more code-oriented scaffolds such as pseudo-code and example code received mixed evaluations, while fill-in-the-blank code was rated least helpful, reflecting skepticism about its pedagogical value. Overall, educators favored scaffolds that provide high-level guidance and reasoning support, rather than more detailed code fragments that risk reducing student engagement in problem solving.

\subsubsection{Boundaries and Control of Scaffolding Levels}

When asked about the maximum level of scaffolding they would allow, educators’ responses varied, though most favored limiting support to example code or fill-in-the-blank code, with fewer supporting step-by-step plans or pseudocode as the upper bound. Their reasoning highlighted the importance of balancing guidance with student autonomy: educators emphasized that “allowing students to write code themselves using examples ensures their mastery of programming” and that overly complete assistance risks “fostering over-reliance.” At the same time, many acknowledged that beginner students need lighter cognitive burdens and more structured supports, making options such as fill-in-the-blank code appropriate for entry-level contexts.

Across these discussions, there was strong agreement that instructors should play a central role in setting boundaries, while students should retain some flexibility within those boundaries. As one educator explained, “instructors may know better what’s good in general, but students want to personalize the setting to some extent to suit their learning styles.” Others stressed the risk of unregulated autonomy, noting that “if students are given full freedom, they will often gravitate toward the maximum help option, undermining deliberate practice.” A minority suggested more dynamic arrangements, such as AI proposing scaffolding levels with student confirmation, but the overall consensus was that instructor-led constraints are essential to align AI use with pedagogical goals.

\subsection{Perceived Benefits, Risks, and Design Considerations}

Educators identified several important advantages of using AI assistants in programming education. They emphasized that AI provides timely and personalized support, enabling students to ask questions freely without fear of judgment, receive guidance outside class hours, and access diverse solutions beyond what instructors might offer. AI was also valued for lowering barriers to practice, sustaining motivation, and helping tailor learning to students’ varied backgrounds.

At the same time, educators consistently highlighted drawbacks and risks, particularly the danger of students bypassing critical reasoning by directly generating answers. Concerns included over-reliance, diminished opportunities to make and learn from mistakes, weakened problem-solving and critical thinking skills, and the possibility of AI outputs being misaligned with course goals or overly advanced for novices.

To better promote high-quality thinking, educators suggested that AI should provide scaffolding rather than final solutions—such as hints, step-by-step prompts, Socratic questioning, and personalized guidance based on student context. They stressed the importance of balancing support and fading to encourage independence, and several proposed that AI should actively ask students questions to stimulate reflection rather than supplying complete answers.

To better promote high-quality thinking, educators stressed that AI should provide scaffolding rather than complete solutions, with scaffolds gradually fading to encourage independence. Looking ahead, they envisioned student-facing features such as adjustable response levels, context-aware feedback, and long-term progress tracking, alongside instructor-facing tools including dashboards to monitor interactions, summaries of student queries, and customizable controls over the scope of AI assistance.

\section{Design Implications for AI-Powered Coding Assistants with Metacognitive Support (RQ4)}

\begin{figure}
   \centering
    \includegraphics[width=0.9\textwidth]{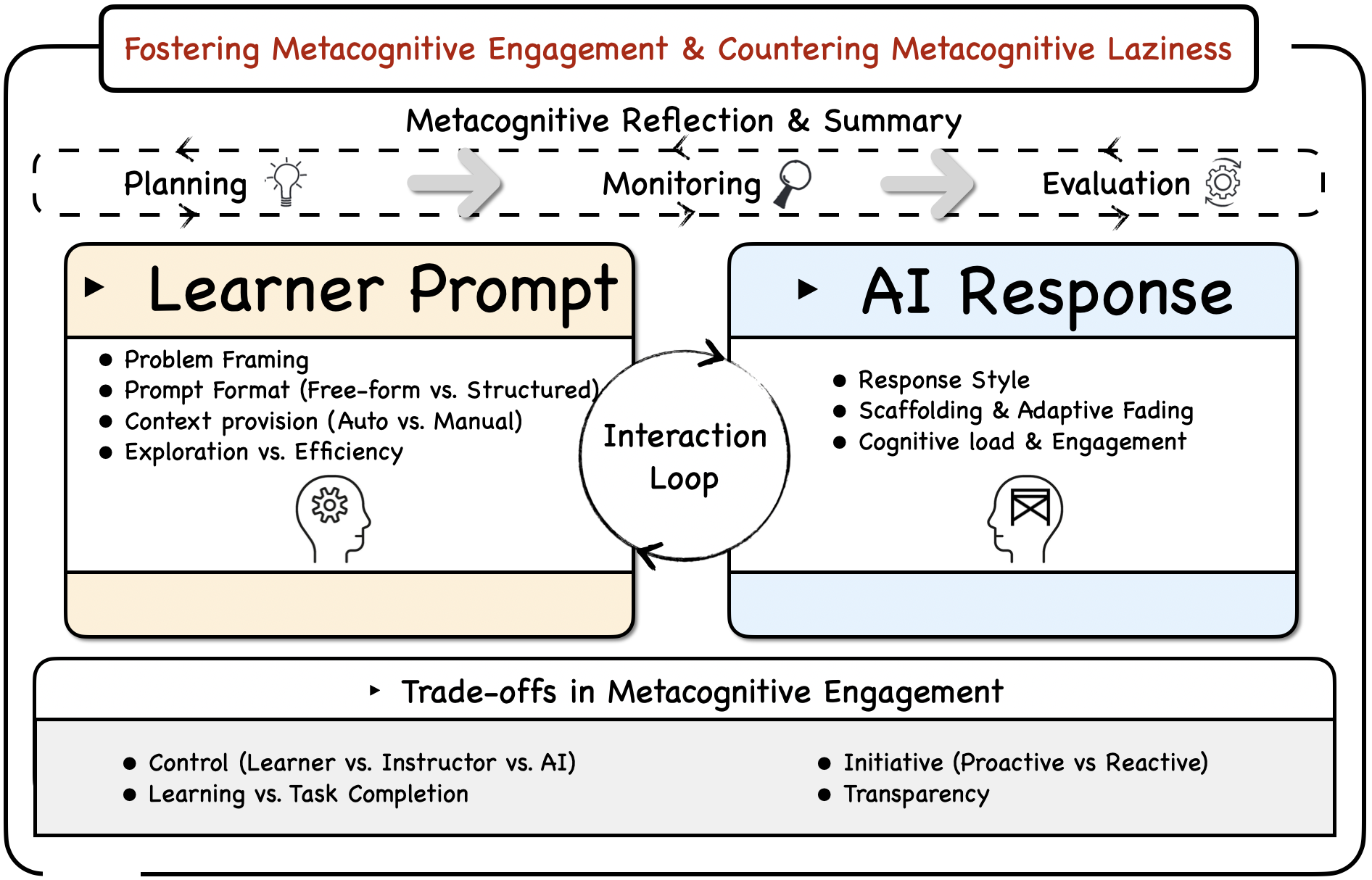}
    \caption{A Conceptual Design Framework for Metacognition-Supportive AI Programming Assistants.}
    \label{design_dimensions}
    \Description{A layered design space linking metacognitive phases (planning–monitoring–evaluation–reflection) with the learner’s prompt formation and the AI’s response scaffolding. Trade-offs span initiative (Proactive vs. Reactive), transparency, and control (learner–instructor–AI), and Learning vs. Task Completion.}
\end{figure}

Drawing on our analysis and survey data from students and educators, we derive design implications for AI-powered coding assistants that improve users’ metacognition through support strategies—for example, scaffolds for planning and evaluation—that can be embedded in GenAI systems. We emphasize that different phases of metacognition entail distinct challenges, risks, and trade-offs, and we highlight opportunities to embed scaffolding that promotes deeper learning rather than superficial task completion. Specifically, as shown in Figure~\ref{design_dimensions}, we structure these implications around student–AI help-seeking interactions: (1) the metacognitive phase of AI engagement, (2) the formulation of a prompt, (3) types and scaffolding of AI responses, and (4) navigating trade-offs in metacognitive engagement.

\subsection{The Metacognitive Phase of AI Engagement}

Our analysis revealed that most student–AI exchanges were confined to a single metacognitive phase, predominantly monitoring, rather than traversing a complete planning-monitoring-evaluation cycle. This fragmentation stems from multiple factors: while some metacognitive activities may occur outside the chat interface (e.g., peer discussions, reviewing notes, consulting learning materials), learners may be overly reliant on AI for immediate assistance rather than engaging in more structured and systematic learning practices \cite{viberg2025chatting}. Moreover, they often lack awareness of the full metacognitive cycle or inadvertently outsource cognitive work to the AI—"metacognitive laziness".

To address this, we propose that metacognition-supportive AI assistants should both cultivate learners’ metacognitive awareness and guide them through complete metacognitive cycles. While recent research has explored explicit phase scaffolding approaches—such as asking learners to select a self-regulated learning phase before querying or providing phase-specific hint categories \cite{li2025coderunner,phung2025plan}—future systems should move toward more integrated conversational designs. These designs would maintain coherence across turns, enabling planning-oriented exchanges to naturally transition into monitoring and ultimately culminate in evaluation phases. In the planning phase, AI can prompt students to explicitly define or restate the problem in their own words, clarify task goals, and specify input–output expectations, thereby fostering early articulation of intent and externalizing tacit assumptions. During monitoring, rather than directly supplying corrections, the system can encourage learners to predict or explain potential errors, compare alternative solutions, or reflect on discrepancies before receiving hints or corrections, ensuring that error interpretation precedes resolution. The evaluation phase should extend beyond correctness checking to broader transfer of learning, such as asking whether a solution generalizes to new tasks, prompting self-explanation or paraphrasing of the final approach, or encouraging optimization by checking boundary cases, efficiency, and readability. Finally, AI assistants could generate reflective summaries of the interaction that highlight key strategies, phases traversed, and moments of difficulty, prompting learners to step back from the task and recognize patterns in their thinking. Strategic use of brief self-explanation prompts can serve as bridges between phases, counteracting metacognitive “laziness” while keeping learners actively engaged throughout the cycle. Together, these design elements situate AI not merely as a problem-solver, but as a partner in metacognitive regulation.

\subsection{The Formulation of a Prompt}

Formulating effective prompts is one of the most critical—and metacognitively demanding—activities in AI-assisted programming. While free-form input, common in chat-based models, feels approachable, our analysis shows that many novices struggle to craft clear prompts. Their inputs are often vague and lack context, leading to unhelpful responses. This reflects a genuine challenge: for beginners, articulating precise questions is not trivial \cite{ma2024exploring}.

Some researchers have argued for systems that automatically gather context—for instance, IDE-style plugins or course-specific AI assistants using retrieval-augmented generation (RAG)—so that models can deliver more well-aligned responses \cite{kazemitabaar2024codeaid}. Yet unlike productivity-oriented programming assistants, educational tools must prioritize supporting learning over simply producing efficient solutions. In this view, structured prompting plays a crucial role. Many educators emphasized that structured prompts foster active engagement, critical thinking, and intentional inquiry. Without scaffolding, however, novices often struggle to frame effective questions and instead fall back on requesting direct answers from AI.

In addition, systems can scaffold prompt crafting through strategies such as task decomposition and prompt chaining—breaking a complex problem into smaller, sequential sub-tasks where outputs from one step inform the next \cite{tankelevitch2024metacognitive}. These strategies not only improve model performance but also help learners clarify goals and refine their reasoning. Complementary feedforward design can further reduce unproductive trial-and-error by signaling when a vague prompt is unlikely to succeed. After the user specifies a task, the system can generate a step-by-step plan for completion, supported by proactive prompting and concrete examples, while still allowing students to skip decomposition to avoid unnecessary cognitive load. For example, when a student requests “write a Python function to sort numbers,” the system might guide them through clarifying prompts such as: What will the input look like (a list of integers, floats, or strings)? What should the output be (a new sorted list, or the same list modified in place)? Should the sorting be ascending or descending? By helping students articulate precise requirements before code is generated, the system transforms prompt formulation into an intentional learning process rather than a shortcut to solutions.

\subsection{Types and Scaffolding of AI Responses}

After formulating a question, the next design consideration is the degree of control over the type and directness of assistance. Both students and educators in our study emphasized the importance of indirect scaffolding, which promotes active engagement rather than passive answer-seeking.

Among common forms of assistance such as hints, step-by-step plans, socratic questioning, pseudo-code, example code, and fill-in-the-blank templates, educators generally preferred hints and step-by-step guidance, as it both reflects learners’ current progress and reinforces best practices. Socratic questioning, as suggested in prior work \cite{kargupta-etal-2024-instruct}, was seen as a way to foster independent problem-solving and encourage critical thinking by transforming code generation requests into a series of reflective prompts. Pseudo-code is already used in some systems, yet several teachers worried it may confuse novices who struggle to bridge abstract logic with executable syntax. Example code remains a crucial learning resource \cite{loksa2016role,kazemitabaar2024codeaid}, but educational AI tools should carefully differentiate between illustrative examples and direct solutions. Fill-in-the-blank code represents another intermediate scaffold, encouraging students to complete missing logic themselves. However, both forms of support are still rarely implemented in current AI assistant tools, and their pedagogical potential warrants further investigation. Importantly, different types of tasks and learner profiles may call for different scaffolding forms. Learners may also exhibit varying preferences depending on factors such as task difficulty, task type, and their stage of learning. Understanding how these dimensions interact remains an open area for future empirical research.

Another dimension concerns the level of help. The key trade-off is providing sufficient scaffolding to promote critical learning while avoiding over-guidance that diminishes autonomy \cite{kazemitabaar2024codeaid}. If support is too direct and learners have complete freedom to choose minimal scaffolding, they may bypass essential learning opportunities within their zone of proximal development, undermining skill growth and self-efficacy. Conversely, overly indirect feedback risks leaving students frustrated and unsupported. To balance these tensions, future systems might adopt adaptive fading strategies.

\subsection{Navigating Trade-offs in Metacognitive Engagement}

Designing AI systems to support metacognitive engagement requires navigating several trade-offs. While such tools can scaffold planning, monitoring, and evaluation, their effectiveness depends on how agency, visibility, and control are distributed across learners, instructors, and the AI itself. Below we discuss three key considerations: who initiates questions, how transparent the process should be, and who ultimately controls the level of assistance.

One important trade-off concerns the locus of initiative in interactions: should the assistant proactively generate questions and prompts to scaffold metacognition, or should it remain reactive and wait for learners to initiate? Proactive designs may reduce metacognitive inertia and help novices who struggle to ask effective questions, while reactive designs preserve learner autonomy and prevent over-direction. A balanced approach may involve adaptive prompting, where the system intervenes only when learners show signs of confusion, stagnation, or superficial engagement.

Transparency is another key factor for sustaining metacognitive engagement. Learners often benefit from visibility into their own progress and the nature of the scaffolding they receive. For example, an AI assistant could provide a monitoring dashboard that tracks which phases of metacognition (planning, monitoring, evaluation) have been engaged, highlights which types of support (hints, examples, explanations) were requested, and visualizes shifts in the level of assistance. Such transparency can help students reflect on their learning patterns, give instructors oversight to adjust pedagogical strategies, and increase trust in the AI system.

Beyond initiative and transparency, a third dimension concerns who controls the level of assistance. Should the AI system make adaptive decisions on its own, should learners be able to fully direct their experience, or should instructors retain authority over the process? Recent work has shown that learners with higher self-efficacy often prefer greater autonomy in choosing their level of support, suggesting that AI learning tools should be context-aware and dynamically adjust the range of control available to learners. In contrast, preferences for the type of help (e.g., hints vs. examples) may be more closely tied to individual style and taste \cite{kargupta-etal-2024-instruct}. A promising direction is to design customizable options that allow learners to calibrate the AI’s support to their needs while still aligning with instructional goals. Future AI assistants could adopt shared and adaptive control rather than relying on fixed settings. Instructors might define the permissible range of support while learners exercise autonomy within those bounds. Such adaptive approaches not only balance autonomy and guidance but also highlight an important design principle: control of scaffolding should be understood as a dynamic, shared process—shaped by learners, instructors, and context—rather than a static parameter.

Taken together, these four dimensions outline a design space for metacognition-supportive AI in programming education. Supporting learners requires attention not only to the phases of metacognitive engagement (planning, monitoring, evaluation), but also to the formulation of prompts that shape how learners articulate their intent, the types and scaffolding of responses that balance directness with inquiry, and the trade-offs in agency, transparency, and control that govern how assistance unfolds in practice. By connecting these dimensions, we emphasize that AI should not be conceived merely as a source of answers, but as a partner that helps learners navigate metacognitive cycles, develop prompting strategies, and engage with adaptive scaffolding. This integrated perspective highlights opportunities for future research to investigate how these layers interact, and how AI can be designed to sustain metacognitive growth across diverse learners and contexts.

\section{Limitations and Future Work}

This study has several limitations. First, our findings derive from a single Python programming course at one institution, which constrains the breadth of applicability. Validation across courses in other languages (such as C or Java), as well as across different institutional contexts and cultural environments, will be important for assessing the robustness of these results. Also, the nature of students’ interactions was closely tied to the capabilities of the LLM models available in each year. The GPT-3.5-turbo model used in 2023 may have constrained the quality of support, shaping student perceptions differently from later cohorts who engaged with GPT-4 and GPT-4o. Beyond model versioning, variation in performance also emerges across different programming tasks. Moreover, this study examined only student–LLM interactions, however students may expressing their critical thinking in ways that are not captured in chatbot logs. In actual classroom contexts, learning unfolds across multiple modalities, such as seeking help from peers or instructors, working within IDEs, consulting learning materials, or referring back to prior exercise solutions. These additional dimensions may play a crucial role in shaping students’ metacognitive processes. Future research should therefore move beyond a single-modal focus by incorporating richer forms of evidence, including classroom video recordings, eye-tracking, and screen-capture data, as well as think-aloud protocols. Such approaches would provide a more comprehensive account of the strategies students adopt and the processes through which they regulate their learning when engaging with AI-powered tools.

\section{Conclusion}

This paper investigated how students engage with AI in programming education, with a particular focus on metacognition processes. By analyzing more than 10,000 dialogue interactions, we identified how students formulated prompts, received feedback, and navigated planning, monitoring, and evaluation processes with AI support. In addition, we complemented this analysis with student surveys that captured learners’ perceptions and experiences, as well as educator surveys that offered insights into the future role of AI-powered tools in programming education. Synthesizing evidence across these sources, we derived high-level design considerations, highlighting key trade-offs that characterize the emerging design space of educational AI tools. We hope that these findings and design implications will help guide the development of future LLM-powered coding assistants that support not only problem solving, but also deeper metacognitive engagement and self-regulated learning.

\bibliographystyle{ACM-Reference-Format}
\bibliography{references}


\begin{thebibliography}{62}


\ifx \showCODEN    \undefined \def \showCODEN     #1{\unskip}     \fi
\ifx \showDOI      \undefined \def \showDOI       #1{#1}\fi
\ifx \showISBNx    \undefined \def \showISBNx     #1{\unskip}     \fi
\ifx \showISBNxiii \undefined \def \showISBNxiii  #1{\unskip}     \fi
\ifx \showISSN     \undefined \def \showISSN      #1{\unskip}     \fi
\ifx \showLCCN     \undefined \def \showLCCN      #1{\unskip}     \fi
\ifx \shownote     \undefined \def \shownote      #1{#1}          \fi
\ifx \showarticletitle \undefined \def \showarticletitle #1{#1}   \fi
\ifx \showURL      \undefined \def \showURL       {\relax}        \fi
\providecommand\bibfield[2]{#2}
\providecommand\bibinfo[2]{#2}
\providecommand\natexlab[1]{#1}
\providecommand\showeprint[2][]{arXiv:#2}

\bibitem[Amani et~al\mbox{.}(2023)]%
        {amani2023generative}
\bibfield{author}{\bibinfo{person}{Sara Amani}, \bibinfo{person}{Lance White},
  \bibinfo{person}{Trini Balart}, \bibinfo{person}{Laksha Arora},
  \bibinfo{person}{Kristi~J Shryock}, \bibinfo{person}{Kelly Brumbelow}, {and}
  \bibinfo{person}{Karan~L Watson}.} \bibinfo{year}{2023}\natexlab{}.
\newblock \showarticletitle{Generative AI perceptions: A survey to measure the
  perceptions of faculty, staff, and students on generative AI tools in
  academia}.
\newblock \bibinfo{journal}{\emph{arXiv preprint arXiv:2304.14415}}
  (\bibinfo{year}{2023}).
\newblock


\bibitem[Anagnostopoulos(2023)]%
        {anagnostopoulos2023chatgpt}
\bibfield{author}{\bibinfo{person}{Christos-Nikolaos Anagnostopoulos}.}
  \bibinfo{year}{2023}\natexlab{}.
\newblock \showarticletitle{ChatGPT impacts in programming education: A recent
  literature overview that debates ChatGTP responses}.
\newblock \bibinfo{journal}{\emph{arXiv preprint arXiv:2309.12348}}
  (\bibinfo{year}{2023}).
\newblock


\bibitem[Becker et~al\mbox{.}(2023)]%
        {becker2023programming}
\bibfield{author}{\bibinfo{person}{Brett~A Becker}, \bibinfo{person}{Paul
  Denny}, \bibinfo{person}{James Finnie-Ansley}, \bibinfo{person}{Andrew
  Luxton-Reilly}, \bibinfo{person}{James Prather}, {and}
  \bibinfo{person}{Eddie~Antonio Santos}.} \bibinfo{year}{2023}\natexlab{}.
\newblock \showarticletitle{Programming Is Hard-Or at Least It Used to Be:
  Educational Opportunities and Challenges of AI Code Generation}. In
  \bibinfo{booktitle}{\emph{Proceedings of the 54th ACM Technical Symposium on
  Computer Science Education V. 1}}. \bibinfo{pages}{500--506}.
\newblock


\bibitem[Bingham and Witkowsky(2021)]%
        {bingham2021deductive}
\bibfield{author}{\bibinfo{person}{Andrea~J Bingham} {and}
  \bibinfo{person}{Patricia Witkowsky}.} \bibinfo{year}{2021}\natexlab{}.
\newblock \showarticletitle{Deductive and inductive approaches to qualitative
  data analysis}.
\newblock \bibinfo{journal}{\emph{Analyzing and interpreting qualitative data:
  After the interview}} (\bibinfo{year}{2021}), \bibinfo{pages}{133--146}.
\newblock


\bibitem[Biswas(2023)]%
        {biswas2023role}
\bibfield{author}{\bibinfo{person}{Som Biswas}.}
  \bibinfo{year}{2023}\natexlab{}.
\newblock \showarticletitle{Role of ChatGPT in Computer Programming.: ChatGPT
  in Computer Programming.}
\newblock \bibinfo{journal}{\emph{Mesopotamian Journal of Computer Science}}
  \bibinfo{volume}{2023} (\bibinfo{year}{2023}), \bibinfo{pages}{8--16}.
\newblock


\bibitem[Brandt et~al\mbox{.}(2009)]%
        {brandt2009two}
\bibfield{author}{\bibinfo{person}{Joel Brandt}, \bibinfo{person}{Philip~J
  Guo}, \bibinfo{person}{Joel Lewenstein}, \bibinfo{person}{Mira Dontcheva},
  {and} \bibinfo{person}{Scott~R Klemmer}.} \bibinfo{year}{2009}\natexlab{}.
\newblock \showarticletitle{Two studies of opportunistic programming:
  interleaving web foraging, learning, and writing code}. In
  \bibinfo{booktitle}{\emph{Proceedings of the SIGCHI Conference on Human
  Factors in Computing Systems}}. \bibinfo{pages}{1589--1598}.
\newblock


\bibitem[Chan and Lee(2023)]%
        {chan2023ai}
\bibfield{author}{\bibinfo{person}{Cecilia Ka~Yuk Chan} {and}
  \bibinfo{person}{Katherine~KW Lee}.} \bibinfo{year}{2023}\natexlab{}.
\newblock \showarticletitle{The AI generation gap: Are Gen Z students more
  interested in adopting generative AI such as ChatGPT in teaching and learning
  than their Gen X and millennial generation teachers?}
\newblock \bibinfo{journal}{\emph{Smart learning environments}}
  \bibinfo{volume}{10}, \bibinfo{number}{1} (\bibinfo{year}{2023}),
  \bibinfo{pages}{60}.
\newblock


\bibitem[Chen et~al\mbox{.}(2025)]%
        {chen2025unpacking}
\bibfield{author}{\bibinfo{person}{Angxuan Chen}, \bibinfo{person}{Mengtong
  Xiang}, \bibinfo{person}{Junyi Zhou}, \bibinfo{person}{Jiyou Jia},
  \bibinfo{person}{Junjie Shang}, \bibinfo{person}{Xinyu Li},
  \bibinfo{person}{Dragan Ga{\v{s}}evi{\'c}}, {and} \bibinfo{person}{Yizhou
  Fan}.} \bibinfo{year}{2025}\natexlab{}.
\newblock \showarticletitle{Unpacking help-seeking process through multimodal
  learning analytics: A comparative study of ChatGPT vs Human expert}.
\newblock \bibinfo{journal}{\emph{Computers \& Education}}
  \bibinfo{volume}{226} (\bibinfo{year}{2025}), \bibinfo{pages}{105198}.
\newblock


\bibitem[Cheng et~al\mbox{.}(2024)]%
        {cheng2024exploring}
\bibfield{author}{\bibinfo{person}{Gary Cheng}, \bibinfo{person}{Di Zou},
  \bibinfo{person}{Haoran Xie}, {and} \bibinfo{person}{Fu~Lee Wang}.}
  \bibinfo{year}{2024}\natexlab{}.
\newblock \showarticletitle{Exploring differences in self-regulated learning
  strategy use between high-and low-performing students in introductory
  programming: An analysis of eye-tracking and retrospective think-aloud data
  from program comprehension}.
\newblock \bibinfo{journal}{\emph{Computers \& Education}}
  \bibinfo{volume}{208} (\bibinfo{year}{2024}), \bibinfo{pages}{104948}.
\newblock


\bibitem[Choi et~al\mbox{.}(2023)]%
        {choi2023benefit}
\bibfield{author}{\bibinfo{person}{Heeryung Choi}, \bibinfo{person}{Jelena
  Jovanovic}, \bibinfo{person}{Oleksandra Poquet}, \bibinfo{person}{Christopher
  Brooks}, \bibinfo{person}{Sre{\'c}ko Joksimovi{\'c}}, {and}
  \bibinfo{person}{Joseph~Jay Williams}.} \bibinfo{year}{2023}\natexlab{}.
\newblock \showarticletitle{The benefit of reflection prompts for encouraging
  learning with hints in an online programming course}.
\newblock \bibinfo{journal}{\emph{The Internet and Higher Education}}
  \bibinfo{volume}{58} (\bibinfo{year}{2023}), \bibinfo{pages}{100903}.
\newblock


\bibitem[Darvishi et~al\mbox{.}(2024)]%
        {darvishi2024impact}
\bibfield{author}{\bibinfo{person}{Ali Darvishi}, \bibinfo{person}{Hassan
  Khosravi}, \bibinfo{person}{Shazia Sadiq}, \bibinfo{person}{Dragan
  Ga{\v{s}}evi{\'c}}, {and} \bibinfo{person}{George Siemens}.}
  \bibinfo{year}{2024}\natexlab{}.
\newblock \showarticletitle{Impact of AI assistance on student agency}.
\newblock \bibinfo{journal}{\emph{Computers \& Education}}
  \bibinfo{volume}{210} (\bibinfo{year}{2024}), \bibinfo{pages}{104967}.
\newblock


\bibitem[Denny et~al\mbox{.}(2023a)]%
        {denny2023conversing}
\bibfield{author}{\bibinfo{person}{Paul Denny}, \bibinfo{person}{Viraj Kumar},
  {and} \bibinfo{person}{Nasser Giacaman}.} \bibinfo{year}{2023}\natexlab{a}.
\newblock \showarticletitle{Conversing with copilot: Exploring prompt
  engineering for solving cs1 problems using natural language}. In
  \bibinfo{booktitle}{\emph{Proceedings of the 54th ACM Technical Symposium on
  Computer Science Education V. 1}}. \bibinfo{pages}{1136--1142}.
\newblock


\bibitem[Denny et~al\mbox{.}(2023b)]%
        {denny2023promptly}
\bibfield{author}{\bibinfo{person}{Paul Denny}, \bibinfo{person}{Juho
  Leinonen}, \bibinfo{person}{James Prather}, \bibinfo{person}{Andrew
  Luxton-Reilly}, \bibinfo{person}{Thezyrie Amarouche},
  \bibinfo{person}{Brett~A Becker}, {and} \bibinfo{person}{Brent~N Reeves}.}
  \bibinfo{year}{2023}\natexlab{b}.
\newblock \showarticletitle{Promptly: Using Prompt Problems to Teach Learners
  How to Effectively Utilize AI Code Generators}.
\newblock \bibinfo{journal}{\emph{arXiv preprint arXiv:2307.16364}}
  (\bibinfo{year}{2023}).
\newblock


\bibitem[Denny et~al\mbox{.}(2023c)]%
        {denny2023computing}
\bibfield{author}{\bibinfo{person}{Paul Denny}, \bibinfo{person}{James
  Prather}, \bibinfo{person}{Brett~A Becker}, \bibinfo{person}{James
  Finnie-Ansley}, \bibinfo{person}{Arto Hellas}, \bibinfo{person}{Juho
  Leinonen}, \bibinfo{person}{Andrew Luxton-Reilly}, \bibinfo{person}{Brent~N
  Reeves}, \bibinfo{person}{Eddie~Antonio Santos}, {and} \bibinfo{person}{Sami
  Sarsa}.} \bibinfo{year}{2023}\natexlab{c}.
\newblock \showarticletitle{Computing Education in the Era of Generative AI}.
\newblock \bibinfo{journal}{\emph{arXiv preprint arXiv:2306.02608}}
  (\bibinfo{year}{2023}).
\newblock


\bibitem[Ebrahimi et~al\mbox{.}(2006)]%
        {ebrahimi2006taxonomy}
\bibfield{author}{\bibinfo{person}{A Ebrahimi}, \bibinfo{person}{D Kopec},
  {and} \bibinfo{person}{C Schweikert}.} \bibinfo{year}{2006}\natexlab{}.
\newblock \showarticletitle{Taxonomy of novice programming error patterns with
  plan, web, and object solutions}.
\newblock \bibinfo{journal}{\emph{Comput. Surveys}} \bibinfo{volume}{38},
  \bibinfo{number}{2} (\bibinfo{year}{2006}), \bibinfo{pages}{1--24}.
\newblock


\bibitem[Fan et~al\mbox{.}(2025)]%
        {fan2025beware}
\bibfield{author}{\bibinfo{person}{Yizhou Fan}, \bibinfo{person}{Luzhen Tang},
  \bibinfo{person}{Huixiao Le}, \bibinfo{person}{Kejie Shen},
  \bibinfo{person}{Shufang Tan}, \bibinfo{person}{Yueying Zhao},
  \bibinfo{person}{Yuan Shen}, \bibinfo{person}{Xinyu Li}, {and}
  \bibinfo{person}{Dragan Ga{\v{s}}evi{\'c}}.} \bibinfo{year}{2025}\natexlab{}.
\newblock \showarticletitle{Beware of metacognitive laziness: Effects of
  generative artificial intelligence on learning motivation, processes, and
  performance}.
\newblock \bibinfo{journal}{\emph{British Journal of Educational Technology}}
  \bibinfo{volume}{56}, \bibinfo{number}{2} (\bibinfo{year}{2025}),
  \bibinfo{pages}{489--530}.
\newblock


\bibitem[Finnie-Ansley et~al\mbox{.}(2022)]%
        {finnie-ansley2022robots}
\bibfield{author}{\bibinfo{person}{James Finnie-Ansley}, \bibinfo{person}{Paul
  Denny}, \bibinfo{person}{Brett~A. Becker}, \bibinfo{person}{Andrew
  Luxton-Reilly}, {and} \bibinfo{person}{James Prather}.}
  \bibinfo{year}{2022}\natexlab{}.
\newblock \showarticletitle{The Robots Are Coming: Exploring the Implications
  of OpenAI Codex on Introductory Programming}. In
  \bibinfo{booktitle}{\emph{Proceedings of the 24th Australasian Computing
  Education Conference}} (Virtual Event, Australia) \emph{(\bibinfo{series}{ACE
  '22})}. \bibinfo{publisher}{Association for Computing Machinery},
  \bibinfo{address}{New York, NY, USA}, \bibinfo{pages}{10–19}.
\newblock
\showISBNx{9781450396431}
\urldef\tempurl%
\url{https://doi.org/10.1145/3511861.3511863}
\showDOI{\tempurl}


\bibitem[Finnie-Ansley et~al\mbox{.}(2023)]%
        {finnie2023my}
\bibfield{author}{\bibinfo{person}{James Finnie-Ansley}, \bibinfo{person}{Paul
  Denny}, \bibinfo{person}{Andrew Luxton-Reilly},
  \bibinfo{person}{Eddie~Antonio Santos}, \bibinfo{person}{James Prather},
  {and} \bibinfo{person}{Brett~A Becker}.} \bibinfo{year}{2023}\natexlab{}.
\newblock \showarticletitle{My AI Wants to Know if This Will Be on the Exam:
  Testing OpenAI’s Codex on CS2 Programming Exercises}. In
  \bibinfo{booktitle}{\emph{Proceedings of the 25th Australasian Computing
  Education Conference}}. \bibinfo{pages}{97--104}.
\newblock


\bibitem[Gao et~al\mbox{.}(2022)]%
        {gao2022who}
\bibfield{author}{\bibinfo{person}{Zhikai Gao}, \bibinfo{person}{Sarah
  Heckman}, {and} \bibinfo{person}{Collin Lynch}.}
  \bibinfo{year}{2022}\natexlab{}.
\newblock \showarticletitle{Who Uses Office Hours? A Comparison of In-Person
  and Virtual Office Hours Utilization}. In
  \bibinfo{booktitle}{\emph{Proceedings of the 53rd ACM Technical Symposium on
  Computer Science Education - Volume 1}} (Providence, RI, USA)
  \emph{(\bibinfo{series}{SIGCSE 2022})}. \bibinfo{publisher}{Association for
  Computing Machinery}, \bibinfo{address}{New York, NY, USA},
  \bibinfo{pages}{300–306}.
\newblock
\showISBNx{9781450390705}
\urldef\tempurl%
\url{https://doi.org/10.1145/3478431.3499334}
\showDOI{\tempurl}


\bibitem[Hao and Liu(2025)]%
        {hao2025towards}
\bibfield{author}{\bibinfo{person}{Qiang Hao} {and} \bibinfo{person}{Ruohan
  Liu}.} \bibinfo{year}{2025}\natexlab{}.
\newblock \showarticletitle{Towards Integrating Behavior-Driven Development in
  Mobile Development: An Experience Report}. In
  \bibinfo{booktitle}{\emph{Proceedings of the 56th ACM Technical Symposium on
  Computer Science Education V. 1}}. \bibinfo{pages}{450--456}.
\newblock


\bibitem[Humble et~al\mbox{.}(2023)]%
        {humble2023cheaters}
\bibfield{author}{\bibinfo{person}{Niklas Humble}, \bibinfo{person}{Jonas
  Boustedt}, \bibinfo{person}{Hanna Holmgren}, \bibinfo{person}{Goran
  Milutinovic}, \bibinfo{person}{Stefan Seipel}, {and}
  \bibinfo{person}{Ann-Sofie {\"O}stberg}.} \bibinfo{year}{2023}\natexlab{}.
\newblock \showarticletitle{Cheaters or AI-Enhanced Learners: Consequences of
  ChatGPT for Programming Education}.
\newblock \bibinfo{journal}{\emph{Electronic Journal of e-Learning}}
  (\bibinfo{year}{2023}), \bibinfo{pages}{00--00}.
\newblock


\bibitem[Ichinco and Kelleher(2015)]%
        {ichinco2015exploring}
\bibfield{author}{\bibinfo{person}{Michelle Ichinco} {and}
  \bibinfo{person}{Caitlin Kelleher}.} \bibinfo{year}{2015}\natexlab{}.
\newblock \showarticletitle{Exploring novice programmer example use}. In
  \bibinfo{booktitle}{\emph{2015 IEEE Symposium on Visual Languages and
  Human-Centric Computing (VL/HCC)}}. IEEE, \bibinfo{pages}{63--71}.
\newblock


\bibitem[Karaoglan~Yilmaz and Yilmaz(2022)]%
        {karaoglan2022learning}
\bibfield{author}{\bibinfo{person}{Fatma~Gizem Karaoglan~Yilmaz} {and}
  \bibinfo{person}{Ramazan Yilmaz}.} \bibinfo{year}{2022}\natexlab{}.
\newblock \showarticletitle{Learning analytics intervention improves
  students’ engagement in online learning}.
\newblock \bibinfo{journal}{\emph{Technology, Knowledge and Learning}}
  \bibinfo{volume}{27}, \bibinfo{number}{2} (\bibinfo{year}{2022}),
  \bibinfo{pages}{449--460}.
\newblock


\bibitem[Kargupta et~al\mbox{.}(2024)]%
        {kargupta-etal-2024-instruct}
\bibfield{author}{\bibinfo{person}{Priyanka Kargupta}, \bibinfo{person}{Ishika
  Agarwal}, \bibinfo{person}{Dilek~Hakkani Tur}, {and} \bibinfo{person}{Jiawei
  Han}.} \bibinfo{year}{2024}\natexlab{}.
\newblock \showarticletitle{Instruct, Not Assist: {LLM}-based Multi-Turn
  Planning and Hierarchical Questioning for Socratic Code Debugging}. In
  \bibinfo{booktitle}{\emph{Findings of the Association for Computational
  Linguistics: EMNLP 2024}}. \bibinfo{pages}{9475--9495}.
\newblock


\bibitem[Kasneci et~al\mbox{.}(2023)]%
        {kasneci2023chatgpt}
\bibfield{author}{\bibinfo{person}{Enkelejda Kasneci}, \bibinfo{person}{Kathrin
  Se{\ss}ler}, \bibinfo{person}{Stefan K{\"u}chemann}, \bibinfo{person}{Maria
  Bannert}, \bibinfo{person}{Daryna Dementieva}, \bibinfo{person}{Frank
  Fischer}, \bibinfo{person}{Urs Gasser}, \bibinfo{person}{Georg Groh},
  \bibinfo{person}{Stephan G{\"u}nnemann}, \bibinfo{person}{Eyke
  H{\"u}llermeier}, {et~al\mbox{.}}} \bibinfo{year}{2023}\natexlab{}.
\newblock \showarticletitle{ChatGPT for good? On opportunities and challenges
  of large language models for education}.
\newblock \bibinfo{journal}{\emph{Learning and Individual Differences}}
  \bibinfo{volume}{103} (\bibinfo{year}{2023}), \bibinfo{pages}{102274}.
\newblock


\bibitem[Kazemitabaar et~al\mbox{.}(2023a)]%
        {kazemitabaar2023studying}
\bibfield{author}{\bibinfo{person}{Majeed Kazemitabaar},
  \bibinfo{person}{Justin Chow}, \bibinfo{person}{Carl Ka~To Ma},
  \bibinfo{person}{Barbara~J Ericson}, \bibinfo{person}{David Weintrop}, {and}
  \bibinfo{person}{Tovi Grossman}.} \bibinfo{year}{2023}\natexlab{a}.
\newblock \showarticletitle{Studying the effect of AI Code Generators on
  Supporting Novice Learners in Introductory Programming}. In
  \bibinfo{booktitle}{\emph{Proceedings of the 2023 CHI Conference on Human
  Factors in Computing Systems}}. \bibinfo{pages}{1--23}.
\newblock


\bibitem[Kazemitabaar et~al\mbox{.}(2023b)]%
        {kazemitabaar2023how_novices_use_llm_code}
\bibfield{author}{\bibinfo{person}{Majeed Kazemitabaar},
  \bibinfo{person}{Xinying Hou}, \bibinfo{person}{Austin Henley},
  \bibinfo{person}{Barbara Ericson}, \bibinfo{person}{David Weintrop}, {and}
  \bibinfo{person}{Tovi Grossman}.} \bibinfo{year}{2023}\natexlab{b}.
\newblock \showarticletitle{How Novices Use LLM-based Code Generators to Solve
  CS1 Coding Tasks in a Self-Paced Learning Environment}. In
  \bibinfo{booktitle}{\emph{Proceedings of the 23rd Koli Calling International
  Conference on Computing Education Research}}.
\newblock


\bibitem[Kazemitabaar et~al\mbox{.}(2024)]%
        {kazemitabaar2024codeaid}
\bibfield{author}{\bibinfo{person}{Majeed Kazemitabaar},
  \bibinfo{person}{Runlong Ye}, \bibinfo{person}{Xiaoning Wang},
  \bibinfo{person}{Austin~Zachary Henley}, \bibinfo{person}{Paul Denny},
  \bibinfo{person}{Michelle Craig}, {and} \bibinfo{person}{Tovi Grossman}.}
  \bibinfo{year}{2024}\natexlab{}.
\newblock \showarticletitle{Codeaid: Evaluating a classroom deployment of an
  llm-based programming assistant that balances student and educator needs}. In
  \bibinfo{booktitle}{\emph{Proceedings of the 2024 chi conference on human
  factors in computing systems}}. \bibinfo{pages}{1--20}.
\newblock


\bibitem[Lau and Guo(2023)]%
        {lau2023ban}
\bibfield{author}{\bibinfo{person}{Sam Lau} {and} \bibinfo{person}{Philip~J
  Guo}.} \bibinfo{year}{2023}\natexlab{}.
\newblock \showarticletitle{From" Ban It Till We Understand It" to" Resistance
  is Futile": How University Programming Instructors Plan to Adapt as More
  Students Use AI Code Generation and Explanation Tools such as ChatGPT and
  GitHub Copilot}. In \bibinfo{booktitle}{\emph{Proceedings of the 2023 ACM
  Conference on International Computing Education Research-Volume 1}}.
\newblock


\bibitem[Leinonen et~al\mbox{.}(2023)]%
        {leinonen2023comparing}
\bibfield{author}{\bibinfo{person}{Juho Leinonen}, \bibinfo{person}{Paul
  Denny}, \bibinfo{person}{Stephen MacNeil}, \bibinfo{person}{Sami Sarsa},
  \bibinfo{person}{Seth Bernstein}, \bibinfo{person}{Joanne Kim},
  \bibinfo{person}{Andrew Tran}, {and} \bibinfo{person}{Arto Hellas}.}
  \bibinfo{year}{2023}\natexlab{}.
\newblock \showarticletitle{Comparing code explanations created by students and
  large language models}. In \bibinfo{booktitle}{\emph{Proceedings of the 2023
  Conference on Innovation and Technology in Computer Science Education V. 1}}.
  \bibinfo{pages}{124--130}.
\newblock


\bibitem[Li and Ma(2025)]%
        {li2025coderunner}
\bibfield{author}{\bibinfo{person}{Huiyong Li} {and} \bibinfo{person}{Boxuan
  Ma}.} \bibinfo{year}{2025}\natexlab{}.
\newblock \showarticletitle{CodeRunner Agent: Integrating AI Feedback and
  Self-Regulated Learning to Support Programming Education}. In
  \bibinfo{booktitle}{\emph{International Conference on Computers in
  Education}}.
\newblock


\bibitem[Liffiton et~al\mbox{.}(2023)]%
        {liffiton2023codehelp}
\bibfield{author}{\bibinfo{person}{Mark Liffiton}, \bibinfo{person}{Brad
  Sheese}, \bibinfo{person}{Jaromir Savelka}, {and} \bibinfo{person}{Paul
  Denny}.} \bibinfo{year}{2023}\natexlab{}.
\newblock \bibinfo{title}{CodeHelp: Using Large Language Models with Guardrails
  for Scalable Support in Programming Classes}.
\newblock
\newblock
\showeprint[arxiv]{2308.06921}~[cs.CY]


\bibitem[Loksa and Ko(2016)]%
        {loksa2016role}
\bibfield{author}{\bibinfo{person}{Dastyni Loksa} {and} \bibinfo{person}{Amy~J
  Ko}.} \bibinfo{year}{2016}\natexlab{}.
\newblock \showarticletitle{The role of self-regulation in programming problem
  solving process and success}. In \bibinfo{booktitle}{\emph{Proceedings of the
  2016 ACM conference on international computing education research}}.
  \bibinfo{pages}{83--91}.
\newblock


\bibitem[Loksa et~al\mbox{.}(2022)]%
        {loksa2022metacognition}
\bibfield{author}{\bibinfo{person}{Dastyni Loksa}, \bibinfo{person}{Lauren
  Margulieux}, \bibinfo{person}{Brett~A Becker}, \bibinfo{person}{Michelle
  Craig}, \bibinfo{person}{Paul Denny}, \bibinfo{person}{Raymond Pettit}, {and}
  \bibinfo{person}{James Prather}.} \bibinfo{year}{2022}\natexlab{}.
\newblock \showarticletitle{Metacognition and self-regulation in programming
  education: Theories and exemplars of use}.
\newblock \bibinfo{journal}{\emph{ACM Transactions on Computing Education
  (TOCE)}} \bibinfo{volume}{22}, \bibinfo{number}{4} (\bibinfo{year}{2022}),
  \bibinfo{pages}{1--31}.
\newblock


\bibitem[Ma et~al\mbox{.}(2024a)]%
        {ma2024enhancing}
\bibfield{author}{\bibinfo{person}{Boxuan Ma}, \bibinfo{person}{Li Chen}, {and}
  \bibinfo{person}{Shin’ichi Konomi}.} \bibinfo{year}{2024}\natexlab{a}.
\newblock \showarticletitle{Enhancing programming education with ChatGPT: a
  case study on student perceptions and interactions in a Python course}. In
  \bibinfo{booktitle}{\emph{International Conference on Artificial Intelligence
  in Education}}. Springer, \bibinfo{pages}{113--126}.
\newblock


\bibitem[Ma et~al\mbox{.}(2024b)]%
        {ma2024exploring}
\bibfield{author}{\bibinfo{person}{Boxuan Ma}, \bibinfo{person}{Li Chen}, {and}
  \bibinfo{person}{Shin’ichi Konomi}.} \bibinfo{year}{2024}\natexlab{b}.
\newblock \showarticletitle{Exploring Student Perception and Interaction using
  CHATGPT in Programming Education}. In \bibinfo{booktitle}{\emph{21st
  International Conference on Cognition and Exploratory Learning in the Digital
  Age, CELDA 2024}}. IADIS Press, \bibinfo{pages}{35--42}.
\newblock


\bibitem[MacNeil et~al\mbox{.}(2023)]%
        {macneil2023experiences}
\bibfield{author}{\bibinfo{person}{Stephen MacNeil}, \bibinfo{person}{Andrew
  Tran}, \bibinfo{person}{Arto Hellas}, \bibinfo{person}{Joanne Kim},
  \bibinfo{person}{Sami Sarsa}, \bibinfo{person}{Paul Denny},
  \bibinfo{person}{Seth Bernstein}, {and} \bibinfo{person}{Juho Leinonen}.}
  \bibinfo{year}{2023}\natexlab{}.
\newblock \showarticletitle{Experiences from using code explanations generated
  by large language models in a web software development e-book}. In
  \bibinfo{booktitle}{\emph{Proceedings of the 54th ACM Technical Symposium on
  Computer Science Education V. 1}}. \bibinfo{pages}{931--937}.
\newblock


\bibitem[Miles and Huberman(1994)]%
        {miles1994qualitative}
\bibfield{author}{\bibinfo{person}{Matthew~B Miles} {and}
  \bibinfo{person}{A~Michael Huberman}.} \bibinfo{year}{1994}\natexlab{}.
\newblock \bibinfo{booktitle}{\emph{Qualitative data analysis: An expanded
  sourcebook}}.
\newblock \bibinfo{publisher}{sage}.
\newblock


\bibitem[National Academies~of Sciences et~al\mbox{.}(2018)]%
        {national2018assessing}
\bibfield{author}{\bibinfo{person}{Engineering National Academies~of Sciences},
  \bibinfo{person}{Medicine}, {et~al\mbox{.}}} \bibinfo{year}{2018}\natexlab{}.
\newblock \bibinfo{booktitle}{\emph{Assessing and responding to the growth of
  computer science undergraduate enrollments}}.
\newblock \bibinfo{publisher}{National Academies Press}.
\newblock


\bibitem[Neuendorf(2017)]%
        {neuendorf2017content}
\bibfield{author}{\bibinfo{person}{Kimberly~A Neuendorf}.}
  \bibinfo{year}{2017}\natexlab{}.
\newblock \bibinfo{booktitle}{\emph{The content analysis guidebook}}.
\newblock \bibinfo{publisher}{sage}.
\newblock


\bibitem[Pankiewicz and Baker(2023)]%
        {pankiewicz2023large}
\bibfield{author}{\bibinfo{person}{Maciej Pankiewicz} {and}
  \bibinfo{person}{Ryan~S Baker}.} \bibinfo{year}{2023}\natexlab{}.
\newblock \showarticletitle{Large Language Models (GPT) for automating feedback
  on programming assignments}.
\newblock \bibinfo{journal}{\emph{arXiv preprint arXiv:2307.00150}}
  (\bibinfo{year}{2023}).
\newblock


\bibitem[Park and Cheon(2025)]%
        {park2025exploring}
\bibfield{author}{\bibinfo{person}{Eunsung Park} {and} \bibinfo{person}{Jongpil
  Cheon}.} \bibinfo{year}{2025}\natexlab{}.
\newblock \showarticletitle{Exploring Debugging Challenges and Strategies Using
  Structural Topic Model: A Comparative Analysis of High and Low-Performing
  Students}.
\newblock \bibinfo{journal}{\emph{Journal of Educational Computing Research}}
  \bibinfo{volume}{62}, \bibinfo{number}{8} (\bibinfo{year}{2025}),
  \bibinfo{pages}{1884--1906}.
\newblock


\bibitem[Phung et~al\mbox{.}(2025)]%
        {phung2025plan}
\bibfield{author}{\bibinfo{person}{Tung Phung}, \bibinfo{person}{Heeryung
  Choi}, \bibinfo{person}{Mengyan Wu}, \bibinfo{person}{Adish Singla}, {and}
  \bibinfo{person}{Christopher Brooks}.} \bibinfo{year}{2025}\natexlab{}.
\newblock \showarticletitle{Plan More, Debug Less: Applying Metacognitive
  Theory to AI-Assisted Programming Education}. In
  \bibinfo{booktitle}{\emph{International Conference on Artificial Intelligence
  in Education}}. Springer, \bibinfo{pages}{3--17}.
\newblock


\bibitem[Phung et~al\mbox{.}(2023)]%
        {phung2023generative}
\bibfield{author}{\bibinfo{person}{Tung Phung},
  \bibinfo{person}{Victor-Alexandru P{\u{a}}durean}, \bibinfo{person}{Jos{\'e}
  Cambronero}, \bibinfo{person}{Sumit Gulwani}, \bibinfo{person}{Tobias Kohn},
  \bibinfo{person}{Rupak Majumdar}, \bibinfo{person}{Adish Singla}, {and}
  \bibinfo{person}{Gustavo Soares}.} \bibinfo{year}{2023}\natexlab{}.
\newblock \showarticletitle{Generative AI for Programming Education:
  Benchmarking ChatGPT, GPT-4, and Human Tutors}.
\newblock \bibinfo{journal}{\emph{International Journal of Management}}
  \bibinfo{volume}{21}, \bibinfo{number}{2} (\bibinfo{year}{2023}),
  \bibinfo{pages}{100790}.
\newblock


\bibitem[Prasad and Sane(2024)]%
        {prasad2024self}
\bibfield{author}{\bibinfo{person}{Prajish Prasad} {and} \bibinfo{person}{Aamod
  Sane}.} \bibinfo{year}{2024}\natexlab{}.
\newblock \showarticletitle{A self-regulated learning framework using
  generative AI and its application in CS educational intervention design}. In
  \bibinfo{booktitle}{\emph{Proceedings of the 55th ACM Technical Symposium on
  Computer Science Education V. 1}}. \bibinfo{pages}{1070--1076}.
\newblock


\bibitem[Prather et~al\mbox{.}(2023)]%
        {prather2023robots}
\bibfield{author}{\bibinfo{person}{James Prather}, \bibinfo{person}{Paul
  Denny}, \bibinfo{person}{Juho Leinonen}, \bibinfo{person}{Brett~A Becker},
  \bibinfo{person}{Ibrahim Albluwi}, \bibinfo{person}{Michelle Craig},
  \bibinfo{person}{Hieke Keuning}, \bibinfo{person}{Natalie Kiesler},
  \bibinfo{person}{Tobias Kohn}, \bibinfo{person}{Andrew Luxton-Reilly},
  {et~al\mbox{.}}} \bibinfo{year}{2023}\natexlab{}.
\newblock \showarticletitle{The robots are here: Navigating the generative ai
  revolution in computing education}.
\newblock In \bibinfo{booktitle}{\emph{Proceedings of the 2023 working group
  reports on innovation and technology in computer science education}}.
  \bibinfo{pages}{108--159}.
\newblock


\bibitem[Rajala et~al\mbox{.}(2023)]%
        {rajala2023call}
\bibfield{author}{\bibinfo{person}{Jaakko Rajala}, \bibinfo{person}{Jenni
  Hukkanen}, \bibinfo{person}{Maria Hartikainen}, {and} \bibinfo{person}{Pia
  Niemel{\"a}}.} \bibinfo{year}{2023}\natexlab{}.
\newblock \showarticletitle{"Call me Kiran" -ChatGPT as a Tutoring Chatbot in a
  Computer Science Course}. In \bibinfo{booktitle}{\emph{Proceedings of the
  26th International Academic Mindtrek Conference}}. \bibinfo{pages}{83--94}.
\newblock


\bibitem[Saliba et~al\mbox{.}(2024)]%
        {saliba2024learning}
\bibfield{author}{\bibinfo{person}{Liam Saliba}, \bibinfo{person}{Elisa
  Shioji}, \bibinfo{person}{Eduardo Oliveira}, \bibinfo{person}{Shaanan
  Cohney}, {and} \bibinfo{person}{Jianzhong Qi}.}
  \bibinfo{year}{2024}\natexlab{}.
\newblock \showarticletitle{Learning with style: Improving student code-style
  through better automated feedback}. In \bibinfo{booktitle}{\emph{Proceedings
  of the 55th ACM Technical Symposium on Computer Science Education V. 1}}.
  \bibinfo{pages}{1175--1181}.
\newblock


\bibitem[Sarsa et~al\mbox{.}(2022)]%
        {sarsa2022automatic}
\bibfield{author}{\bibinfo{person}{Sami Sarsa}, \bibinfo{person}{Paul Denny},
  \bibinfo{person}{Arto Hellas}, {and} \bibinfo{person}{Juho Leinonen}.}
  \bibinfo{year}{2022}\natexlab{}.
\newblock \showarticletitle{Automatic generation of programming exercises and
  code explanations using large language models}. In
  \bibinfo{booktitle}{\emph{Proceedings of the 2022 ACM Conference on
  International Computing Education Research-Volume 1}}.
  \bibinfo{pages}{27--43}.
\newblock


\bibitem[Savelka et~al\mbox{.}(2023)]%
        {savelka2023thrilled}
\bibfield{author}{\bibinfo{person}{Jaromir Savelka}, \bibinfo{person}{Arav
  Agarwal}, \bibinfo{person}{Marshall An}, \bibinfo{person}{Chris Bogart},
  {and} \bibinfo{person}{Majd Sakr}.} \bibinfo{year}{2023}\natexlab{}.
\newblock \showarticletitle{Thrilled by Your Progress! Large Language Models
  (GPT-4) No Longer Struggle to Pass Assessments in Higher Education
  Programming Courses}. In \bibinfo{booktitle}{\emph{Proceedings of the 2023
  ACM Conference on International Computing Education Research-Volume 1}}.
\newblock


\bibitem[Shoufan(2023)]%
        {shoufan2023exploring}
\bibfield{author}{\bibinfo{person}{Abdulhadi Shoufan}.}
  \bibinfo{year}{2023}\natexlab{}.
\newblock \showarticletitle{Exploring Students’ Perceptions of CHATGPT:
  Thematic Analysis and Follow-Up Survey}.
\newblock \bibinfo{journal}{\emph{IEEE Access}} (\bibinfo{year}{2023}).
\newblock


\bibitem[Silva et~al\mbox{.}(2024)]%
        {silva2024learning}
\bibfield{author}{\bibinfo{person}{Leonardo Silva},
  \bibinfo{person}{Ant{\'o}nio Mendes}, \bibinfo{person}{Anabela Gomes}, {and}
  \bibinfo{person}{Gabriel Fortes}.} \bibinfo{year}{2024}\natexlab{}.
\newblock \showarticletitle{What Learning Strategies are Used by Programming
  Students? A Qualitative Study Grounded on the Self-regulation of Learning
  Theory}.
\newblock \bibinfo{journal}{\emph{ACM Transactions on Computing Education}}
  \bibinfo{volume}{24}, \bibinfo{number}{1} (\bibinfo{year}{2024}),
  \bibinfo{pages}{1--26}.
\newblock


\bibitem[Skjuve et~al\mbox{.}(2023)]%
        {skjuve2023user}
\bibfield{author}{\bibinfo{person}{Marita Skjuve}, \bibinfo{person}{Asbj{\o}rn
  F{\o}lstad}, {and} \bibinfo{person}{Petter~Bae Brandtzaeg}.}
  \bibinfo{year}{2023}\natexlab{}.
\newblock \showarticletitle{The user experience of ChatGPT: Findings from a
  questionnaire study of early users}. In \bibinfo{booktitle}{\emph{Proceedings
  of the 5th International Conference on Conversational User Interfaces}}.
  \bibinfo{pages}{1--10}.
\newblock


\bibitem[Smith et~al\mbox{.}(2017a)]%
        {smith2017my}
\bibfield{author}{\bibinfo{person}{Aaron~J. Smith},
  \bibinfo{person}{Kristy~Elizabeth Boyer}, \bibinfo{person}{Jeffrey Forbes},
  \bibinfo{person}{Sarah Heckman}, {and} \bibinfo{person}{Ketan Mayer-Patel}.}
  \bibinfo{year}{2017}\natexlab{a}.
\newblock \showarticletitle{My Digital Hand: A Tool for Scaling Up One-to-One
  Peer Teaching in Support of Computer Science Learning}. In
  \bibinfo{booktitle}{\emph{Proceedings of the 2017 ACM SIGCSE Technical
  Symposium on Computer Science Education}} (Seattle, Washington, USA)
  \emph{(\bibinfo{series}{SIGCSE '17})}. \bibinfo{publisher}{Association for
  Computing Machinery}, \bibinfo{address}{New York, NY, USA},
  \bibinfo{pages}{549–554}.
\newblock
\showISBNx{9781450346986}
\urldef\tempurl%
\url{https://doi.org/10.1145/3017680.3017800}
\showDOI{\tempurl}


\bibitem[Smith et~al\mbox{.}(2017b)]%
        {smith2017office}
\bibfield{author}{\bibinfo{person}{Margaret Smith}, \bibinfo{person}{Yujie
  Chen}, \bibinfo{person}{Rachel Berndtson}, \bibinfo{person}{Kristen~M
  Burson}, {and} \bibinfo{person}{Whitney Griffin}.}
  \bibinfo{year}{2017}\natexlab{b}.
\newblock \showarticletitle{"Office Hours Are Kind of Weird": Reclaiming a
  Resource to Foster Student-Faculty Interaction.}
\newblock \bibinfo{journal}{\emph{InSight: A Journal of Scholarly Teaching}}
  \bibinfo{volume}{12} (\bibinfo{year}{2017}), \bibinfo{pages}{14--29}.
\newblock


\bibitem[Sun et~al\mbox{.}(2024)]%
        {sun2024would}
\bibfield{author}{\bibinfo{person}{Dan Sun}, \bibinfo{person}{Azzeddine
  Boudouaia}, \bibinfo{person}{Chengcong Zhu}, {and} \bibinfo{person}{Yan Li}.}
  \bibinfo{year}{2024}\natexlab{}.
\newblock \showarticletitle{Would ChatGPT-facilitated programming mode impact
  college students’ programming behaviors, performances, and perceptions? An
  empirical study}.
\newblock \bibinfo{journal}{\emph{International Journal of Educational
  Technology in Higher Education}} \bibinfo{volume}{21}, \bibinfo{number}{1}
  (\bibinfo{year}{2024}), \bibinfo{pages}{14}.
\newblock


\bibitem[Tankelevitch et~al\mbox{.}(2024)]%
        {tankelevitch2024metacognitive}
\bibfield{author}{\bibinfo{person}{Lev Tankelevitch}, \bibinfo{person}{Viktor
  Kewenig}, \bibinfo{person}{Auste Simkute}, \bibinfo{person}{Ava~Elizabeth
  Scott}, \bibinfo{person}{Advait Sarkar}, \bibinfo{person}{Abigail Sellen},
  {and} \bibinfo{person}{Sean Rintel}.} \bibinfo{year}{2024}\natexlab{}.
\newblock \showarticletitle{The metacognitive demands and opportunities of
  generative AI}. In \bibinfo{booktitle}{\emph{Proceedings of the 2024 CHI
  Conference on Human Factors in Computing Systems}}. \bibinfo{pages}{1--24}.
\newblock


\bibitem[Tian et~al\mbox{.}({[n.\,d.]})]%
        {tian2304chatgpt}
\bibfield{author}{\bibinfo{person}{H Tian}, \bibinfo{person}{W Lu},
  \bibinfo{person}{TO Li}, \bibinfo{person}{X Tang}, \bibinfo{person}{SC
  Cheung}, \bibinfo{person}{J Klein}, {and} \bibinfo{person}{TF
  Bissyand{\'e}}.} \bibinfo{year}{[n.\,d.]}\natexlab{}.
\newblock \showarticletitle{Is ChatGPT the ultimate programming assistant—how
  far is it?(2023)}.
\newblock \bibinfo{journal}{\emph{arXiv preprint arXiv:2304.11938}}
  (\bibinfo{year}{[n.\,d.]}).
\newblock


\bibitem[Tlili et~al\mbox{.}(2023)]%
        {tlili2023if}
\bibfield{author}{\bibinfo{person}{Ahmed Tlili}, \bibinfo{person}{Boulus
  Shehata}, \bibinfo{person}{Michael~Agyemang Adarkwah}, \bibinfo{person}{Aras
  Bozkurt}, \bibinfo{person}{Daniel~T Hickey}, \bibinfo{person}{Ronghuai
  Huang}, {and} \bibinfo{person}{Brighter Agyemang}.}
  \bibinfo{year}{2023}\natexlab{}.
\newblock \showarticletitle{What if the devil is my guardian angel: ChatGPT as
  a case study of using chatbots in education}.
\newblock \bibinfo{journal}{\emph{Smart Learning Environments}}
  \bibinfo{volume}{10}, \bibinfo{number}{1} (\bibinfo{year}{2023}),
  \bibinfo{pages}{15}.
\newblock


\bibitem[Viberg et~al\mbox{.}(2025)]%
        {viberg2025chatting}
\bibfield{author}{\bibinfo{person}{Olga Viberg}, \bibinfo{person}{Jacqueline
  Wong}, \bibinfo{person}{Yael Feldman-Maggor}, \bibinfo{person}{Nora Dunder},
  {and} \bibinfo{person}{Carrie~Demmans Epp}.} \bibinfo{year}{2025}\natexlab{}.
\newblock \showarticletitle{Chatting with Code: Exploring LLMs as Learning
  Partners in Programming Education}. In
  \bibinfo{booktitle}{\emph{International Conference on Artificial Intelligence
  in Education}}. Springer, \bibinfo{pages}{453--461}.
\newblock


\bibitem[Yilmaz and Yilmaz(2023)]%
        {yilmaz2023augmented}
\bibfield{author}{\bibinfo{person}{Ramazan Yilmaz} {and} \bibinfo{person}{Fatma
  Gizem~Karaoglan Yilmaz}.} \bibinfo{year}{2023}\natexlab{}.
\newblock \showarticletitle{Augmented intelligence in programming learning:
  Examining student views on the use of ChatGPT for programming learning}.
\newblock \bibinfo{journal}{\emph{Computers in Human Behavior: Artificial
  Humans}} \bibinfo{volume}{1}, \bibinfo{number}{2} (\bibinfo{year}{2023}),
  \bibinfo{pages}{100005}.
\newblock


\bibitem[Zhang et~al\mbox{.}(2024)]%
        {zhang2024students}
\bibfield{author}{\bibinfo{person}{Zhengdong Zhang}, \bibinfo{person}{Zihan
  Dong}, \bibinfo{person}{Yang Shi}, \bibinfo{person}{Thomas Price},
  \bibinfo{person}{Noboru Matsuda}, {and} \bibinfo{person}{Dongkuan Xu}.}
  \bibinfo{year}{2024}\natexlab{}.
\newblock \showarticletitle{Students’ perceptions and preferences of
  generative artificial intelligence feedback for programming}. In
  \bibinfo{booktitle}{\emph{Proceedings of the AAAI Conference on Artificial
  Intelligence}}, Vol.~\bibinfo{volume}{38}. \bibinfo{pages}{23250--23258}.
\newblock


\end{thebibliography}

\newpage
\appendix

\onecolumn

\section{Thematic Analysis Codebook}~\label{appendix_codebook}
This appendix includes the final codebook used for the thematic analysis of students' usage of AI.

\renewcommand{\thetable}{A1}

\begin{table*}[h]
\centering
\caption{Thematic Analysis Codebook: \textbf{Prompt} - \textit{What do students ask, and how do these requests reflect their metacognitive processes?}}
\small
\renewcommand{\arraystretch}{1.2}
\begin{tabularx}{\textwidth}{>{\hsize=0.01\hsize}l>{\hsize=0.34\hsize}X>{\hsize=0.65\hsize}X}
\specialrule{.1em}{.05em}{.05em} 
& \textbf{Phase and Prompt Type} & \textbf{Code Description (Strategy)} \\ \hline
&\textbf{Planning} & \textbf{Students seek support in designing initial solution strategies} \\
P1 & Conceptual Question (CQ) & Ask questions about programming-related conceptual knowledge \\
P2 & Problem Understanding (PU)  & Ask the purpose of the question \\
P3 & Asking for Example (AE) & Request example code for specific concepts or functionalities \\
P4 & Code Implementation Question (CI) & Ask specific questions related to code implementation \\
P5 & Code Generation (CG) & Request the generation of code snippets for specific requirements\\
\specialrule{.1em}{.05em}{.05em} 
&\textbf{Monitoring} & \textbf{Students seek support in identifying and resolving issues during code execution} \\
M1 & Error Message Interpretation (EM)  & Request interpretation of the meaning and causes of error messages \\
M2 & Code Correction (CC) & Ask for debugging and correction of errors in the code \\
M3 & Code Verification (CV) & Ask for verification of the correctness of code \\
\specialrule{.1em}{.05em}{.05em} 
&\textbf{Evaluation} & \textbf{Students seek support in reflecting on and improving code quality} \\
E1 & Code Explanation (CE) & Ask for explanation of the functionality and purpose of the code \\
E2 & Code Optimization (CO) & Request the improvement of existing code based on specific needs \\
  \specialrule{.1em}{.05em}{.05em} 

\end{tabularx}
\Description{Table showing a section of the thematic analysis codebook. Each code is accompanied by its respective description.}
\end{table*}

\renewcommand{\thetable}{A2}

\begin{table*}[h]
\centering
\small
\caption{Thematic Analysis Codebook: \textbf{Response} - How much is the response directly revealing the solution?}
\renewcommand{\arraystretch}{1.2}
\begin{tabularx}{\textwidth}{>{\hsize=0.01\hsize}l>{\hsize=0.34\hsize}X>{\hsize=0.65\hsize}X}
\specialrule{.1em}{.05em}{.05em} 
& \textbf{Dimensions and Codes} & \textbf{Code Description} \\ \hline
& \textbf{How does the response reveal the solution?} & \parbox[t]{\hsize}{\textit{How much is the response directly revealing the solution?}} \\ 
S1 & Exact Solution Code & Generated the code solution to a question \\
S2 & Step to Fix Semantic Issue & Generated the steps required to fix semantic/logical problems, which usually need additional lines to achieve new functionality \\ 
S3 & Step to Fix Syntax Issue & Generated the steps required to fix minor syntax issues\\ 
S4 & Step to Fix External Issue & Generated the steps to fix an issue that is not within the code, but about the compilation or execution \\ 
S5 & Example Code & Generated a generic example for a function, or a generic implementation \\
S6 & Conceptual Explanation &Provides explanations and clarifications of programming concepts\\ 
S7 & Code Explanation & Provides explanations and commentary on specific code snippets\\ 
\specialrule{.1em}{.05em}{.05em} 
 &\textbf{How technically correct?} & 
  \textit{Despite the question, how correct is the response from AI?} \\ 
 & Correct & Everything including the answer and its explanation is correct \\ 
 & Incorrect & Any part of the answer or explanation is incorrect \\ 
\specialrule{.1em}{.05em}{.05em} 
& \textbf{How helpful?} & \textit{Is the response helpful to students? Does it guide them the right direction based on the provided query? Does it identify their potential issues? or is it completely misleading?} \\ 
 & Helpful & Answer that allows the student to take one step further, even if it is not arriving at the final solution\\ 
 & Not Helpful & Answer that does not allow students to progress any further, or it is unrelated \\ 
\specialrule{.1em}{.05em}{.05em} 
\end{tabularx}
\end{table*}

\section{Student Perception Questionnaire}~\label{appendix_student}
This appendix lists the pre- and post-questionnaire items assessing students’ perceptions of GenAI for programming learning.

\renewcommand{\thetable}{B1}

\begin{table}[h]
\small
\setlength{\tabcolsep}{2mm}{
\centering
\renewcommand{\arraystretch}{1.2} 
\caption{Pre-questionnaire items.}
\begin{tabular}{cl} \hline
\textbf{Questions} & \textbf{Item} \\ \hline
Pre1 &What is your current level of proficiency in Python programming?\\
& \textbullet\ Beginner  \textbullet\ Intermediate  \textbullet\ Advanced \\
Pre2 &Do you currently use GenAI (e.g., ChatGPT)?\\
& \textbullet\ I use them regularly.\\
& \textbullet\ I used to use them but not anymore.\\
& \textbullet\ I haven't used them but I plan to.\\
& \textbullet\ I don't use them and have no plans to.\\
Pre3& How familiar are you with GenAI (e.g., ChatGPT)?\\
& \textbullet\ I can explain them in detail.\\
& \textbullet\ I can explain them to some extent.\\
& \textbullet\ I have a basic understanding.\\
& \textbullet\ I have little understanding.\\
& \textbullet\ I have no understanding at all.\\
Pre4 &To what extent do you think using GenAI (e.g., ChatGPT) to learn Python programming is beneficial or detrimental?\\
& \textbullet\ I think it is positive.\\
& \textbullet\ I rather think it is positive.\\
& \textbullet\ I am neutral.\\
& \textbullet\ I rather think it is negative.\\
& \textbullet\ I think it is negative.\\
Pre5 & In your own words, how would you describe GenAI such as ChatGPT? (Open-ended)\\ \hline
\end{tabular}}
\end{table}

\renewcommand{\thetable}{B2}

\begin{table}[h]
\centering
\caption{Post-questionnaire items.}
\small
{
\setlength{\tabcolsep}{2mm} 
\renewcommand{\arraystretch}{1.2}
\begin{tabular}{cl} \hline
\textbf{Questions} & \textbf{Item} \\ \hline
Post1 & To what extent do you think GenAI (e.g., ChatGPT) are helpful for learning programming? (5-point Likert)\\
Post2 & To what extent are you likely to continue using GenAI (e.g., ChatGPT) for learning programming? (5-point Likert)\\
Post3 & To what extent are you likely to use GenAI (e.g., ChatGPT) frequently in your studies? (5-point Likert)\\
Post4 & To what extent are you likely to recommend GenAI (e.g., ChatGPT) to your peers? (5-point Likert)\\
Post5 & How do you think GenAI (e.g., ChatGPT) can support programming learning? (Multiple answers)\\
& \textbullet\ Correct programming code.\\
& \textbullet\ Answer programming questions.\\
& \textbullet\ Provide examples of programming code.\\
& \textbullet\ Offer learning advice and resources.\\
& \textbullet\ Explain programming concepts.\\
Post6 & In your own words, how would you describe GenAI such as ChatGPT after actually using them? (Open-ended)\\
Post7 & What do you see as the advantages of using GenAI (e.g., ChatGPT) for learning programming? (Open-ended)\\
Post8 & What do you see as the disadvantages of using GenAI (e.g., ChatGPT) for learning programming? (Open-ended)\\
Post9 & In your view, how could GenAI (e.g., ChatGPT) be improved to better support programming learning? (Open-ended)\\
\hline
\end{tabular}
} 
\end{table}

\section{Educator Survey}~\label{appendix_educator}
This appendix lists the educator-survey items.

\renewcommand{\thetable}{C}

\begin{table}[h]
\setlength{\tabcolsep}{2mm}
\centering
\small
\caption{Educator survey items.}
\renewcommand{\arraystretch}{1.2}
\begin{tabularx}{\linewidth}{L{25mm}Y}
\toprule
\textbf{Questions} & \textbf{Item} \\
\midrule
\multicolumn{2}{l}{\textbf{A. Background}}\\
\addlinespace[2pt]
\texttt{A1} & Teaching experience \\
  & \textbullet\  0-1 years \textbullet\ 1-3 years \textbullet\ 3-5 years \textbullet\ 5-10 years \textbullet\ 10-20 years \textbullet\ Over 20 years \\
\texttt{A2} & Teaching level \\
            & \textbullet\  Undergraduate  \textbullet\  Graduate \\
\texttt{A3} & Course type \\
            & \textbullet\  Intro to Programming \textbullet\ Data Science / Machine Learning \\
\texttt{A4} & Programming languages used in course \\
            & \textbullet\ Python \textbullet\ C/C++  \textbullet\ Java \\
\texttt{A5} & Current AI policy in your course \\
            & \textbullet\ Prohibited  \textbullet\ Allowed with Conditions   \textbullet\ Mostly Open \\
\texttt{A6} & Briefly describe your current AI policy and reason in your course (Open-ended) \\
\texttt{A7} & Overall, AI assistants help students learn programming in my course (5-point Likert) \\
\addlinespace[4pt]
\midrule
\multicolumn{2}{l}{\textbf{B. Input \& Context}}\\
\addlinespace[2pt]
\texttt{B1} & Please rate the following student question types by perceived learning value \\
                & \textbullet\ Conceptual Question (5-point Likert)  \\
                & \textbullet\ Asking for Advice / Learning Resources (5-point Likert)  \\
                & \textbullet\ Problem Understanding (5-point Likert) \\
                & \textbullet\ Asking for Example (5-point Likert) \\
                & \textbullet\ Code Implementation Questions (5-point Likert) \\
                & \textbullet\ Code Generation (5-point Likert) \\
                & \textbullet\ Error Message Interpretation (5-point Likert) \\
                & \textbullet\ Code Correction (5-point Likert) \\
                & \textbullet\ Code Verification (5-point Likert) \\
                & \textbullet\ Code Explanation (5-point Likert) \\
                & \textbullet\ Code Optimization (5-point Likert) \\
\texttt{B2} & When students ask for help, which input format better promotes high-quality thinking? \\
               & \textbullet\ Structured Form (specific structure or format designed by educators/experts) \\
               & \textbullet\ Free-form (students can ask questions in a free form manner) \\
\texttt{B3} & Please briefly explain why you chose this option. (Open-ended) \\
\texttt{B4} & Should the system automatically gather course-specific context, or require manual input from students? \\
              & \textbullet\ Automatically gather context \textbullet\ Manual input from students \\
\texttt{B5} & Please briefly explain why you chose this option. (Open-ended) \\
\addlinespace[4pt]
\midrule
\end{tabularx}
\end{table}

\begin{table}[t]
\ContinuedFloat
\setlength{\tabcolsep}{2mm}
\centering
\small
\caption[]{Educator survey items (continued).} 
\renewcommand{\arraystretch}{1.2}
\begin{tabularx}{\linewidth}{L{25mm}Y}
\toprule
\textbf{Questions} & \textbf{Item} \\
\midrule
\multicolumn{2}{l}{\textbf{C. Scaffolding \& Delivery}}\\
\addlinespace[2pt]
\texttt{C1} & How the system’s response should be delivered? \\
               & \textbullet\ Direct solutions (e.g., runnable code or full fix) \\
               & \textbullet\ Indirect scaffolding (e.g., hints/steps/pseudocode, error interpretation) \\
\texttt{C2} & Please rate the following indirect scaffolding by perceived helpfulness\\
               & \textbullet\ Hints / Clues (5-point Likert) \\
               & \textbullet\ Step-by-step Plan (5-point Likert) \\
               & \textbullet\ Socratic Questions (5-point Likert) \\
               & \textbullet\ Pseudo-code (5-point Likert) \\
               & \textbullet\ Example Code (5-point Likert) \\
               & \textbullet\ Fill-in-the-Blank Code (5-point Likert) \\
\texttt{C3} & Select the maximum indirect scaffolding style you would allow \\
            & \textbullet\  No AI assistance \\
            & \textbullet\  Only Hints/Clues  \\
            & \textbullet\ Up to Step-by-step plan  \\
            & \textbullet\ Up to Pseudo-code  \\
      	    & \textbullet\ Up to Example code  \\
	    & \textbullet\ Up to Fill in the Blank Code  \\
	    & \textbullet\ Up to Direct runnable code  \\
      
\texttt{C4} & Please briefly explain why you chose this option (Open-ended) \\
\texttt{C5} & Who should control the answer’s scaffolding style? \\
             & \textbullet\ Instructor controls all  \\
	    & \textbullet\ Instructor sets a maximum, students choose within range  \\
	    & \textbullet\ Student chooses freely  \\
	    & \textbullet\ System suggests and student confirms  \\
	    & \textbullet\ System controls all \\
\texttt{C6} & Please briefly explain why you chose this option (Open-ended) \\
\addlinespace[4pt]
\midrule
\multicolumn{2}{l}{\textbf{D. Pros, Cons, and Feature Wishes}}\\
\addlinespace[2pt]
\texttt{D1} & What are the most important advantages of students using AI assistants for programming learning in your course? (Open-ended) \\
\texttt{D2} & What are the most important drawbacks or risks of students using AI assistants in your course? (Open-ended) \\
\texttt{D3} & What kind of features of the system do you want to better promote students' high-quality thinking? (Open-ended) \\
\texttt{D4} & From your teaching perspective, what features would you expect the system to provide? (Open-ended) \\
\bottomrule
\end{tabularx}
\end{table}

\end{document}